\documentclass[12pt,preprint]{aastex}
\usepackage{graphicx}
\usepackage{natbib}
\usepackage{latexsym}

\bibpunct{(}{)}{;}{a}{}{,}

\shorttitle{Metal Content of the high-z IGM} 
\shortauthors{Frank et al.}

\begin{document} 

\title{A Survey of Metal Lines at High-redshift (I) : SDSS Absorption Line Studies - The Methodology and First Search Results for OVI}

\author{S. Frank\altaffilmark{1},
        S. Mathur\altaffilmark{1},
        M. Pieri\altaffilmark{1},
           \and D. G. York\altaffilmark{2}}

\altaffiltext{1}{Department of Astronomy, Ohio State University, 140 W.18th Ave., Columbus, OH 43210, USA}
\altaffiltext{2}{Department of Astronomy and Astrophysics, University of Chicago, 5640 S.Ellis Avenue, Chicago, Illinois 6063x7, USA}
\email{frank@astronomy.ohio-state.edu}

\begin{abstract} 
We report the results of a systematic search for signatures of metal lines in quasar spectra of the Sloan Digital Sky Survey (SDSS) Data Release 3(DR3), focusing on finding intervening absorbers via detection of their \ion{O}{6}{}  doublet. Here we present the search algorithm, and criteria for distinguishing candidates from spurious Lyman $\alpha${} forest lines. In addition, we compare our findings with simulations of the Lyman $\alpha${} forest in order to estimate the detectability of \ion{O}{6}{}  doublets over various redshift intervals. We have obtained a sample of 1756 \ion{O}{6}{}  doublet candidates with rest-frame equivalent width $\geq${} 0.05 \AA{} in 855 AGN spectra (out of 3702 objects with redshifts in the accessible range for \ion{O}{6}{}  detection). This sample is further subdivided into 3 groups according to the likelihood of being real and the potential for follow-up observation of the candidate. The group with the cleanest and most secure candidates is comprised of 145 candidates. 69 of these reside at a velocity separation $\geq${} 5000 km/s from the QSO, and can therefore be classified tentatively as intervening absorbers. Most of these absorbers have not been picked up by earlier, automated QSO absorption line detection algorithms. This sample increases the number of known \ion{O}{6}{} absorbers at redshifts beyond z$_{abs} \geq${} 2.7 substantially.
\end{abstract}


\section{Introduction}\label{introduction}
Our understanding of the nature of the intergalactic medium (IGM) and its evolution as traced by the absorption features observed in the sightlines towards luminous objects like quasars, has benefitted tremendously within the past few years both from observational and theoretical advances. High-resolution studies of the Lyman $\alpha${} forest have been extended to very low column densities due to high-resolution echelle spectrographs on powerful 8m class telescopes like HIRES (Keck) and UVES (VLT) \citep[]{hu1995, lu1996, kim1997, kirkman1997}. At the same time, theoretical models incorporating gas dynamics, radiative cooling, and photoionisation, have been developed that can reproduce the majority of the observed properties of the quasar absorption spectra \citep[]{cen1994, miraldaescude1996, hernquist1996, petitjean1995, dave1997}. In this picture, baryonic gas can be encountered in a wide variety of physical conditions. Within the Lyman $\alpha${} forest, at least at high redshift where the forest is the main repository for baryons, most of the gas resides in relatively cool (T$\sim 10^4${} K) low-to-medium overdensity structures that are not in dynamical or thermal equilibrium and are mostly governed by photoionisation.\\
The IGM is expected to be highly ionised with a neutral fraction of hydrogen f(\ion{H}{1}) $\sim 10^{-4} $ to $10^{-6}$ . In such a medium, metals are also highly ionised. Thus, oxygen, as the most abundant intergalactic metal, may exist in the \ion{O}{6}{} state. Since the \ion{O}{6} 1032/1038 \AA{} doublet is observable from ground-based telescopes for redshifts beyond z$_{abs} \sim${} 2.0, these transitions constitute a primary tool for studying the characteristics of the metals in the IGM at high redshift. The density of the IGM is sufficiently low as to allow the production of \ion{O}{6}{} by photoionisation from the intergalactic UV and soft X-ray radiation field \citep[]{haardt1996, schaye2000, simcoe2004}. Indeed, \ion{O}{6}{} absorbers seen towards higher redshifts can usually be modelled by structures that are photoionised \citep[]{bergeron2002, carswell2002, levshakov2003}.\\
Due to the different wavebands required for detections at different redshifts and the strong evolution of the Lyman forest, a variety of different techniques and observational strategies have to be applied for finding and identifying \ion{O}{6}{} absorbers. Detection of the \ion{O}{6}{} 1032/1038\AA{} doublet from ground based telescopes is only possible beyond redshifts of $z_{lim}{}  \geq$2.0 due to the 3000 \AA{} atmospheric cutoff, and thus studies of the absorber statistics below this threshold need to rely on space-based instruments. \citet[]{burles1996} conducted the first systematic survey of \ion{O}{6}{} absorbers at z $\sim${} 1.0, employing FOS on HST to study a sample of 11 QSOs, and found 12 suitable candidates of which 9 were expected to be real, thus establishing that the number density per redshift interval of these absorbers is similar to, if not greater than, that for \ion{C}{4}{} and \ion{Mg}{2}{} absorbers at the same redshift.\\
Towards higher redshifts, despite the advantange of shifting the lines into the range for ground-based instruments, the ever increasing density of the Lyman forest renders unambigious identifications of \ion{O}{6}{} absorbers in its midst more difficult than detections made longward of the Lyman $\alpha${} forest. Thus, most studies have relied upon high-resolution and high-signal-to-noise ratio spectra. Around redshifts of z $\sim${} 2.0-3.0, a variety of surveys find evidence for a population of absorbers residing in low to medium overdense regions, mostly photoionised by the ambient extragalactic radiation field \citep[]{carswell2002, bergeron2002, aracil2004, simcoe2004}. The metal enrichment of the IGM inferred by theses surveys ranges from $10^{-2}${} to $10^{-3}${} of the solar value for structures around the mean density of the universe at $z \sim${} 2.0. Using UVES spectra for 2 lensed quasar systems, \citet[]{lopez2007} find size constraints for such intervening \ion{O}{6}{} absorbers with lower boundaries on the kpc scale, where no or only very little variation in different lines of sights are seen. This is broadly consistent with other studies of binary QSO sightline inferring correlation lengths of the absorbers up to several tens of kpc \citep[]{smette1995, petitjean1998, lopez2000, rauch2001}.\\
An interesting method to statistically infer the \ion{C}{4}{} abundance from measuring the mean \ion{C}{4}{} optical depth associated with all pixels of the Lyman $\alpha${} forest with similar optical depths, was pioneered by \citet[]{cowie1998}. The authors find a general correlation of $\tau (CIV)${} with $\tau (HI)${} when $\tau (HI) \geq${} 1.0. This method was applied to \ion{O}{6}{} by \citet[]{songaila1998}, \citet[]{schaye2000} and recently \citet[]{aracil2005} who assert the existence of \ion{O}{6}{} in gas where the \ion{H}{1}{} optical depth is as low as 0.1. There is, however, some ambiguity in the interpretation of the results of such statistical pixel-to-pixel correlations. For an extensive discussion of the method see \citet[]{aguirre2002}, and \citet[]{pieri2004} for a critical analysis of the results.\\
The mechanism for enriching the IGM with metals to the level inferred by these surveys remains somewhat unclear, yet there are at least good candidates : early and wide-spread dissemination of metals by Population III star formation on pre-galactic structures at very high redshifts \citep[]{nath1997, ferrara2000, barkana2001, madau2001} or winds and superwinds from starbursts within galaxies at later stages \citep[]{aguirre2001, adelberger2003}. For the latter mechanism, driving these winds out of the dense environments into the low density IGM remains a crucial point of contention, and a variety of methods have been suggested ranging from supernovae \citep[]{couchman1986}, to ejection by mergers \citep[]{gnedin1997}, or photoevaporation \citep[]{barkana2001}.\\
Table 1 gives an overview of different surveys for direct \ion{O}{6}{} detection in a variety of environments and over a wide redshift range. \citet[]{reimers2006}{} find 6 different \ion{O}{6}{} systems and a further 8 potential, but blended,  candidates in a single object (HS0747+4259) at 1.46 $\leq z_{abs} \leq ${} 1.81, deriving a redshift path density of $dN_{OVI}/dX \sim$ 13. \citet[]{simcoe2004}{} analyse 230 Lyman $\alpha${} forest lines with a hydrogen column density N$_{HI} \geq 10^{13.6}${} cm$^{-2}$, for 7 QSOs with $z_{em} \sim$ 2.6, and retrieve a total of about 50 accompanying \ion{O}{6}{} systems with secure identifications. Using the capabilities of VLT and UVES within the 'Large Programme : The Cosmic Evolution of the IGM', \citet[]{bergeron2005}{} survey 10 bright QSO at 2.1 $\leq z_{em} \leq${} 2.8, and find 136 \ion{O}{6}{} candidates with 12.7 $\leq log N_{OVI} \leq${} 14.6 in 51 systems. \citet[]{carswell2002} focus on two z$_{em}${} = 2 QSOs and hydrogen absorbers with log N$_{HI} \geq${} 14.0. They identify 7 individual \ion{O}{6}{} lines in 2  intervening systems in one case, and find 13 individual lines in 10 intervening systems in the other. \citet[]{fox2007} study the frequency of \ion{O}{6}{} absorbers associated with 35 damped and sub-damped Lyman $\alpha${} systems, and report 12 detections of which 9 are intervening. \citet[]{lopez2007}{} present 10 intervening \ion{O}{6}{} absorbers seen towards two lensed QSO pairs at a median redshift of z$_{abs}${} = 2.3. Our sample, residing at z$_{abs} \geq 2.7$, therefore greatly increases the redshift range of known \ion{O}{6}{} absorbers at high redshifts. It will allow us to test expectation based upon photoionisation models like the ones presented in \citet[]{dave1998}, thereby possibly constraining the physical conditions of the IGM and the metagalactic UV/X-ray background at high-redshifts.
\\
In this paper, we have decided to apply a direct pixel-by-pixel search for signatures of strong \ion{O}{6}{} 1032/1038 \AA{} features seen in the spectra of SDSS QSOs. This is, to our knowledge, the first systematic survey at high redshifts (2.7 $\leq z_{em} \leq$5.0) at low resolution (R$\sim$1800) and low signal-to-noise ratio. We demonstrate that there is a redshift window of opportunity where the density of the Lyman forest is not yet too high to obliterate the hope for finding and identifying relatively narrow metal absorbers, and that the pathlength covered by the combined SDSS sample is high enough to expect finding a reasonable number of such absorbers given conservative assumptions about the metallicity of the IGM, the abundance of \ion{O}{6}{} due to photoionisation and the average signal-to-noise ratio of the QSO data sample.\\
What type of \ion{O}{6}{} absorbers do we expect to detect via such a direct search without a priori selecting on other transitions ? If a sightline passes through a galaxy, and if there are analogues to local examples surveyed early by \citet[]{rogerson1973, york1974}{} and \citet[]{york1977}{} for highly ionised metals in absorption, we anticipate to detect in such cases \ion{O}{6}{} mixed with \ion{C}{4}{} in the  gas phase inside that galaxy, like in the disk of our Galaxy. Furthermore, a recent, thorough examination of FUSE spectra \citep[]{wakker2003} revealed the occurence of local \ion{O}{6}{} absorbers in almost all of the more than 100 sightlines probed by extragalactic background objects, indicating a large covering fraction when a sightline passes the Galaxy. While the majority of these absorbers are probably located within the Galaxy \citep[]{savage2003}, especially the high-velocity  \ion{O}{6}{} (relative to the Local Standard of Rest) traces a variety of different environments and phenomena, most likely including warm/hot gas interaction in an extended Galactic corona and even truly intergalactic gas within the Local Group \citep[]{sembach2003}. As an aside, the gas residing in the background AGN host galaxy is subject to a strong UV radiation field by the central engine itself, and thus might contain more \ion{O}{6}{} than normal galaxies. And while we are per se not interested in such absorbers for the study of intergalactic gas, we have included 'intrinsic' absorbers, as they are being found with our algorithm as well.
\\
From hydrodynamic simulations \citet[]{cen2001}{} assert, however, that the majority of detectable lines with rest frame equivalent widths EW$_{r}(OVI 1032 \AA) \geq 0.035 \AA${} do {\it not} reside in virialised regions like galaxies, groups and clusters, but trace intergalactic gas at only moderate overdensities ($\sim$10-40). Although their estimated redshift path density of about 5 such absorbers per redshift of unity drops rapidly by a factor of at least ten when increasing the equivalent width limit to 0.35 \AA, even that lower rate of incidence allows us to detect an appreciable number of such strong intergalactic \ion{O}{6}{} absorbers, owing to the large redshift path length probed by the collective SDSS AGN sample.
\\
To summarise, we expect to detect the signals of \ion{O}{6}{} absorbers in a variety of environments. While this first step of finding \ion{O}{6}{} rests upon a 'blind' \ion{O}{6}{} doublet search plus an alignment with Lyman $\alpha${} and $\beta${} lines, in order to ensure high purity of the candidate sample, we may then, in a future effort, use independent line detections of different ions  to better characterise the physical stae and potentially the nature of the absorbers found here. Because there are a variety of possible origins of the OVI gas, including  outflows, intervening galaxies, and the contribution of the AGN host galaxy, this next step is needed to arrive at a sample of IGM absorbers. There is evidence for a population of \ion{O}{6}{} absorbers that contain little, if any detectable hydrogen, as pointed out by \citet[]{bergeron2005}, and it is clear that our search algorithm presented here cannot detect those. However, as we anticipate the need for high-resolution follow-up of our candidates in order to confirm their nature, we might then be able to learn about this oxygen-rich subset (type 0 of \citet[]{bergeron2005}) of absorbers as well.
\\
The paper is organised as follows : section \ref{prelim}{} details our preliminary analyses of whether the signal-to-noise ratio of the SDSS spectra and the density of the Lyman forest enables our search program. In section \ref{agn_sample}{} we present our SDSS QSO sample selection, and section \ref{search_strategy}{} gives an overview of the search algorithms we applied. Before we summarise and conclude in section \ref{summary}, we present the results of our search in section \ref{candidate_list}. The search method developed here can be easly tailored to a variety of different ionic species. We stress that follow-up observations of the candidates we retrieve with high S/N and high resolution is neccessary to determine the reliability of our search algorithm by determining the nature of the features we detect.
\\
Throughout this study we use a cosmology with H$_{0}$=71 km s$^{-1}$Mpc$^{-1}$, $\Omega _{M}$ = 0.27 and $\Omega _{\lambda}$=0.73. Abundances are given by number relative to hydrogen, and vacuum wavelength are being used, if not otherwise noted.           


\section{Preliminary feasibility analyses}\label{prelim}
With rest-frame wavelengths below the \ion{H}{1}{}  Lyman $\alpha$ wavelength of 1215.67 \AA{}, the \ion{O}{6}{}  1032\AA/1038\AA{} doublet absorption lines inevitably fall into the Lyman alpha forest of systems located along the same sightline with slightly lower redshifts. Thus, two effects may prohibit our attempts to find such \ion{O}{6}{}  absorbers : first, the ubiquity of the Ly$\alpha${} forest lines, especially at higher redshifts, can lead to blending of the oxygen lines with \ion{H}{1}{}  lines, thus rendering a correct identification of the doublet impossible. Second, two different Ly $\alpha${} lines at different redshifts and the accidental occurence of optical depth ratios expected for \ion{O}{6}{} can mimic an \ion{O}{6}{}  doublet, leading to a false identification.
\subsection{The expected number of falsely identified absorbers}
In order to estimate the severity of the latter effect, we have created mock catalogues of Lyman forests lines, using the line density estimate of \citet{kim2001} and the column density distribution function of \citet{hu1995} :
\begin{equation}\label{line_density}
\frac{dn}{dz} = 9.06\times(1+z)^{2.19\pm0.27} \textrm{ for 13.64 }< \textrm{ log N}_{HI} < 16.0
\end{equation}
\begin{equation}\label{cddf}
f(N_{HI})dN_{HI} = 4.9 \times 10^{7} \times N_{HI}^{-1.46} \times dN_{HI}
\end{equation}
where $\frac{dn}{dz}${} denotes the number of Lyman $\alpha${} lines per redshift interval dz, and f(N$_{HI}${}) is the probability distribution of obtaining a line with column density of neutral hydrogen N$_{HI}${} within the interval N$_{HI}$, N$_{HI}$+dN$_{HI}$.
We then analysed these catalogs, and retrieved all line pairs that exhibit the correct wavelength ratio (within a given tolerance) as well as the correct ratio of optical depths at the line center  :
\begin{equation}\label{wave_ratio}
r_{1} = \frac{\lambda _{OVI (1032\AA)}}{\lambda _{OVI (1038\AA)}} = 0.99452
\end{equation}
\begin{equation}\label{EW_ratio}
r_{2} = \frac{\tau (OVI (1032\AA))}{\tau (OVI (1038\AA))} = 2.0
\end{equation}
$r_{1}${} is independent of redshift since both lines get shifted by the same factor (1+z), and the result for $r_{2}${} follows from atomic physics :
\begin{equation}\label{optical_depth}
\tau_{0} = N_{j}\frac{\lambda ^{4}}{8\pi^{3/2}c}\frac{g_{k}}{g_{j}}\frac{A_{kj}}{b_{\lambda}}
\end{equation}
where N is the column density, $\lambda$ the wavelength of the transition, c the speed of light, A$_{kj}$ the Einstein-coefficient for the transition from the upper (k) to the lower (j) level, b$_\lambda =\lambda_{0}\frac{b}{c}$ the line broadening factor. The statistical weights $g_{k}/g_{j}${} of the two transitions is equal to two.
\\
With the given resolution of SDSS spectra (R$\sim$1800) and the conservative assumption that line centroids can be determined to about 1/3 of the resolution element, we expect a tolerance level for the wavelength ratio of $\frac{\Delta r1}{r1}$ = 3.0$\times$10$^{-4}$. Furthermore, we can rule out all pairs that do not exhibit the range of flux decrements, expected from the ratios of the two line strengths convolved with the instrument capabilities. The efficiency of this procedure depends crucially on the ability to precisely measure the equivalent width (EW) of lines, but even a conservative estimate of a 20\%{} accuracy on a single pixel flux measurement, corresponding to a low signal-to-noise ratio of about 5,  leads to a reduction of the number of false detections of roughly a factor of eight at all redshifts.\\
An analysis of the mock data sets spanning the complete redshift range available to us (2.7 $< z_{abs} < z_{max}$(SDSS)) and varying the thresholds on the desired precision for $r_{1}${} and $r_{2}${} yields for the number of expected interlopers:
\begin{equation}\label{interlopers}
n_{i}(z_{abs}) = n_{i,0}(z_{abs}) (\frac{\Delta r_{1}/r_{1}}{3.0\times10^{-4}}) (\frac{\Delta r_{2}/r_{2}}{0.2})^{1.12}
\end{equation}
where $z_{abs}${} is the absorber redshift, and $n_{i}${} is the number of expected absorbers per spectrum. n$_{i,0} (z_{abs})${} is a strongly increasing function of the redshift:
\begin{equation}\label{g_z}
n_{i,0}(z_{abs}) = 2.75\times10^{-4} (1+z_{abs})^{7.5}
\end{equation}
Note that the coefficients and exponents in equations \ref{interlopers}{} and \ref{g_z}{} are derived from analysing the mock data set, without taking additional absorption from metal lines into account. Nor did we analyse the contribution of random noise features mimicking \ion{O}{6}{} doublets. The latter problem, however, is obviated by the fact that we are going to focus on lines that are clearly above the noise limit, as we will explain in section 3.\\
We focus on \ion{O}{6}{} absorbers accompanied by strong \ion{H}{1}{}  Lyman absorption\footnote{Note that Bergeron et al. (2002), Carswell et al. (2002), and Bergeron \& Herbert-Fort (2005) report some unusual \ion{O}{6}{} absorbers with high oxygen abundances (-1 $<$ [O/H] $<$ 0). Such absorbers, classified by Bergeron \& Herbert-Fort as type 0, could potentially slip through our search criterion as they need not be accompanied by strong \ion{H}{1}{} absorption. A priori we cannot determine which fraction of these absorbers we might lose, some of them exhibit strong enough \ion{H}{1}{} absorption to pass our criteria (right panel Fig. 2 of Bergeron \& Herbert-Fort (2005)).}. Therefore, we rule out all lines that do not show accompanying \ion{H}{1}{}  transitions. Given the line density estimate of \citet{kim2001}, the lower limit for the average velocity separation of two Lyman $\alpha${} absorbers is 
\begin{equation}\label{del_v}
\Delta v _{Ly \alpha} = c \times \frac{\Delta z}{1+z_{abs}} = c \times \frac{\Delta n}{9.06 \times (1+z_{em})^{2.19} \times (1+z_{abs})} \geq 30,000 \textrm{ km/s } \times (1+z_{em})^{-3.19}
\end{equation}
where we used the relation $z_{abs} \leq z_{em}${} to obtain the lower limit in the last step, and equated $\Delta n = 1$.\footnote{Note that we do not take line clustering into account here, which leads to a decreased velocity separation in areas of high line density.}\\
The velocity separation of one pixel in SDSS spectra, however, amounts to 
\begin{equation}\label{del_v_SDSS}
\Delta v _{SDSS} = c \times \frac{\Delta \lambda _{pixel}}{\lambda} \sim 70 \textrm{ km/s.}
\end{equation}
Thus, the probability to have at least one \ion{H}{1}{}  line fall right onto the pixel corresponding to the redshift of an ``\ion{O}{6}{}  pixel'' is 
\begin{equation}\label{prop_interloper}
p _{interloper} = \frac{\Delta v _{SDSS}}{\Delta v _{Ly \alpha}} \sim 2.3\times 10^{-3} \times (1+z_{em})^{3.19}
\end{equation}
Note that the \citet[]{kim2001}{} line density includes Lyman $\alpha${} absorbers with column densities as low as log N$_{HI}${} = 13.64. Restricting ourselves to absorbers that show at least an optical depth at line centre for the Lyman $\beta${} line of $\tau _{0}$(Lyman $\beta$) = 1.0, leads to another factor of f$_{strong} \sim${} 0.4 in the exclusion of spurious interlopers.
Therefore, we can further reduce the number of spurious interlopers by $p_{interloper} \times f_{strong}$, at least, by requiring the candidate \ion{O}{6}{} feature to be accompanied by strong Lyman $\alpha${} and $\beta${}{} absorption. Since higher order series lines become rather weak, adding additional components probably does not follow the same simple multiplicative propagation of probabilities, and so we have considered in this feasibility study only the first two terms of the Lyman series.  
\\

\begin{figure}
\includegraphics[angle=270,width=\columnwidth]{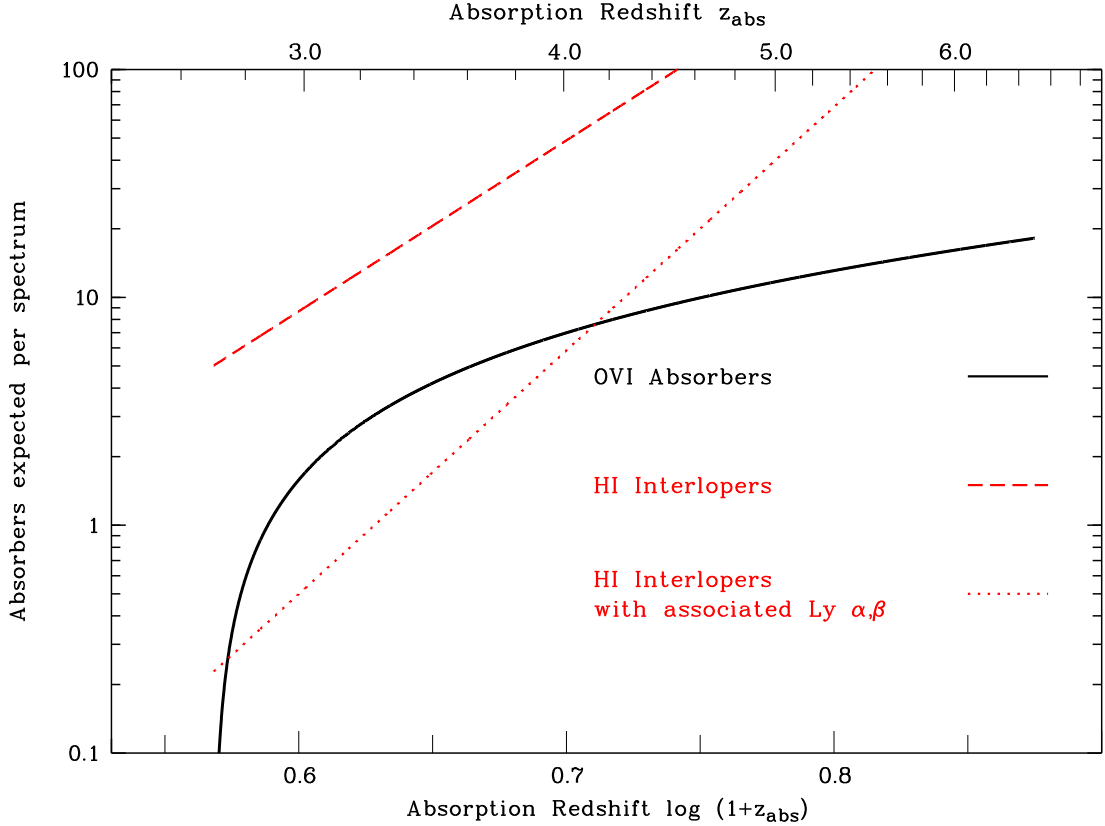}\label{expected_lines}\caption{The number of \ion{O}{6}{}  absorption features and spurious \ion{H}{1}{}  interlopers expected in the spectra of SDSS quasars versus the redshift z$_{abs}${} of the absorber. The rapid increase for the \ion{O}{6}{}  lines up to log (1+z$_{abs})\sim 0.65$ (=z$_{abs} \sim${} 3.4) can mainly be attributed to the nearly linearly growing redshift pathlength, whereas the rapid increase of \ion{H}{1}{}  interlopers is due to the redshift evolution of the line density. We assume an [O/H] = -2.0 to calculate the number of absorbers by integrating over the column density distribution by Hu et al. (1995), after assessing a correction factor for the fraction of lines retrievable (for details see section 2.2). The red, dashed curve represents the expected number of \ion{H}{1}{}  interlopers that exhibit less than a fiducial value for the deviations in both line position and line strength (for details see text), whereas the dotted line takes the additional requirement into account that the interlopers have to show associated, strong \ion{H}{1}{}  absorption. It is obvious that our search efficiency peaks around a redshift of z$_{abs} \sim${} 3.2, and rapidly deteriorates for redshifts z$_{abs} \geq 4.0$. Thus, we expect our ''window of opportunity'' to be between 2.8 $< z_{abs} < $ 4.0, where the ratio of detections of real systems to spurious interlopers is greater than unity.}
\end{figure}

Figure 1 shows in red the number of expected \ion{H}{1}{}  interlopers without (dashed) or with (dotted) the additional requirement of accompanying Lyman $\alpha${} and $\beta${} absorption. For these calculations we have used the fiducial values of $\frac{\Delta r_{1}}{r_{1}} \leq 3\times 10^{-4}$ and $\frac{\Delta r_{2}}{r_{2}} \leq 0.2$. Introducing the additional requirement  significantly reduces the number of false positives, as Figure 1 demonstrates.\\

Finally,  we can also disqualify a potential candidate as belonging to an \ion{O}{6}{} doublet if it is clear that it is part of a Lyman series itself, i.e. we can check whether there are lower (or also higher) Lyman transitions at the same redshift with similar velocity structures. This cut becomes possible for all lines above 4500 \AA{} when the Lyman $\beta${} line redshifts into the lower wavelength regime of SDSS. Thus, we expect to even further reduce the number of falsely classified \ion{O}{6}{}  absorbers beyond a redshift of z$_{OVI} \sim${} 3.37 by a large fraction. Another criterion that can be used to correctly identify a metal line doublet is the shape of the lines : both are expected to have the same velocity structure and should be rather narrow, in stark contrast to the Lyman lines that tend to be broader than 20 km/s. Due to the low resolution of SDSS spectra, however, this criterion cannot be implemented here unless the \ion{H}{1}{} line is so strong that its width exceeds one resolution element. Such strongly saturated or even damped lines are rare, and would have been picked up by other automatic search algorithms already, as we describe later.\\

We will discuss the implementation and success of these and various other criteria to distinguish real \ion{O}{6}{}  lines from \ion{H}{1}{}  features mimicking metal lines in section 3.  

\subsection{The detectability of \ion{O}{6}{}  doublets in the Ly $\alpha${} forest}
In order to estimate whether blending and/or noise renders the identification of existing \ion{O}{6}{}  doublets in the Lyman forest impossible, we have created 100 mock spectra of a typical Lyman forest, following the number and column density distributions from above, and adjusting the spectral resolution and pixel size to the SDSS fiducial value. For the Doppler broadening parameter, we have assumed a Gaussian distribution of width 8 km/s around a mean of $b = 25 km/s$, the exact values for the distribution are, however, not important at this stage of modelling, as the instrumental resolution broadens the line profiles over more than the average velocity width. In this forest we placed \ion{O}{6}{} doublets at a redshift of about 3.5, added noise according to the appropriate (worst-case) SDSS scenario (i.e. we have chosen a value of S/N = 3 here, which is below the nominal S/N $>$ 4 SDSS limit, see section 3.3 for details of the SDSS sample's S/N), and checked carefully how often we could reliably retrieve the \ion{O}{6}{}  doublet, when varying the strength of the absorption feature (until it completely vanished in the noise).  Figure 2 shows two examples : in the first, both components are clearly identified, whereas in the second, both are severely blended. Of course, stronger lines can be identified more easily. From this analyis, we conclude that at least 15\%{} of systems with an EW = log (EW$_{\lambda 1032\AA, obs}) = -0.8${} can be unambigiously identified at a redshift of z $\sim${} 3.5, and  hence higher fractions at lower z or greater strength. Because the modelling at this stage is only suppposed to render an initial estimate of the line detectability, we have not included second-order effects like line clustering, or a detailed description of the effects of continuum placement errors.\\

\begin{figure}
\includegraphics[angle=270,width=\columnwidth]{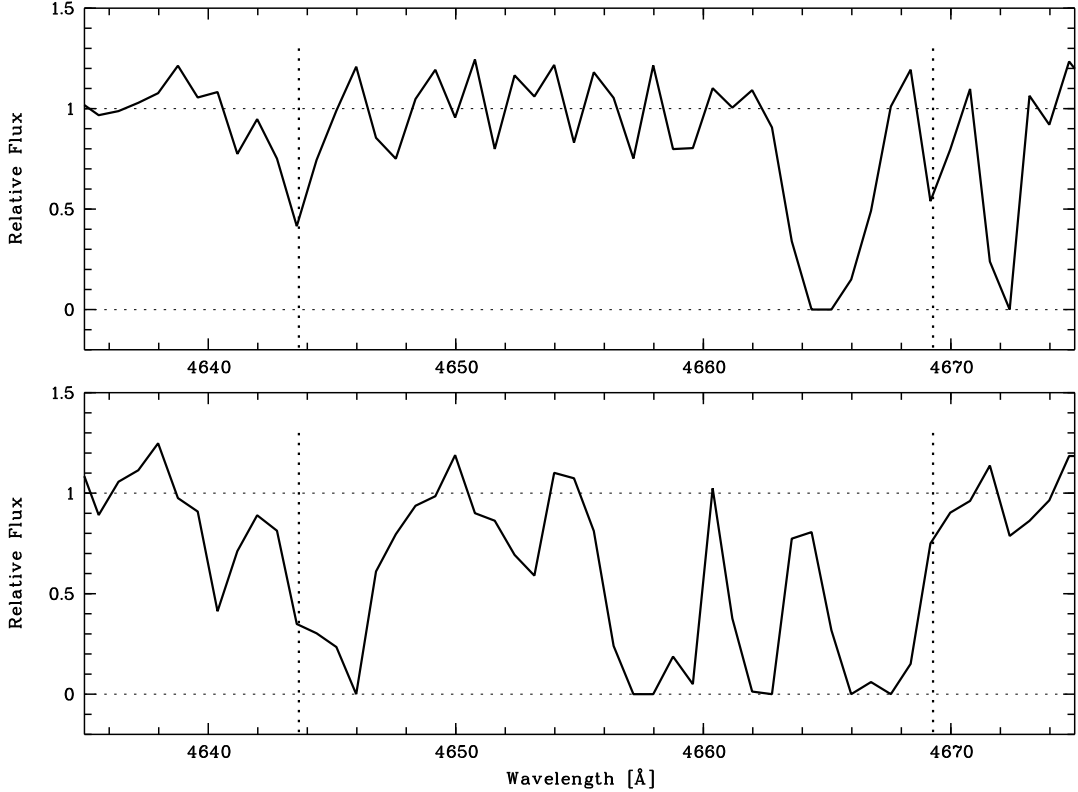}\label{mock_spectrum1}\caption{Two examples of mock normalised AGN spectra with a resolution typical for the SDSS set-up. On top of the Lyman $\alpha${} forest features, generated according to the column density distribution and line densities discussed in the text, an \ion{O}{6}{}  doublet with EW$_{\lambda, obs}$ = 0.15 \AA{} was placed at redshift z$_{abs} \sim$ 3.5. The positions of the two lines are indicated by the dotted line. Then the worst case level of noise for the SDSS spectra was added. {\bf Upper panel:} In this example both absorption lines can be easily identified and they maintain the expected  optical depth ratio.{\bf Lower panel:} In this example, both components are blended with Lyman forest lines, changing the optical depth ratio drastically. Note also how noise affected the 1038 \AA{} component and shifted its flux upwards. Clearly, such a scenario renders the identification of the doublet impossible.}
\end{figure}

How many such absorbers do we expect per sightline ? The above limit for the EW lies only slightly higher than the Line Observability Index (LOX\footnote{The Line of Sight Observability Index(LOX) is defined properly in \citet[]{hellsten1998}. It predicts a rest equivalent width $W_{\lambda r}${} for an absorption line of a given species by modelling gas irradiated with a given X-ray/UV background with CLOUDY \citep[]{cloudy}. For our purposes, LOX $\sim${} log $(W_{\lambda r}/1 m\AA)$} , cf. \citet{hellsten1998}{}) derived by \citet{dave1998}{} for a metallicity of [O/H]$= - 2.0$\footnote{Here, and in the following, the bracket notation is used to indicate the logarithm of the oxygen to hydrogen abundance relative to the sun : [O/H] = log [N(O)/N(H)] - log [N(O)/N(H)]$\odot$}. Integrating the \citet{hu1995}{} distribution over the interval $13.8 <$ log N$_{HI} < 15.4$, where the LOX is of order 1.5 and above (cf. fig.2 of \citet{dave1998}), and along the sightline to a QSO to a redshift z$_{abs}$ of the absorber, we obtain the number n(z$_{abs}$) of such lines associated with detectable \ion{O}{6}{} absorption: 

\begin{displaymath}\label{N_zabs}
n(z_{abs}) = \int_{13.8}^{15.4} \int_{2.7}^{z_{em}} f(N_{HI}) r(LOX, z_{em}) \textrm{dlog } N_{HI} \frac{dX}{dz} dz 
\end{displaymath} 
where $\frac{dX}{dz} = (\frac{(1+z)^{2}}{(1+z)^{2}(1+\Omega_{M}z)-z(z+2)\Omega_{\Lambda}})^{\frac{1}{2}}$ is the redshift path length, and r(LOX, z$_{em}$) is the estimate for the fraction of lines correctly identifiable at the emission redshift of the QSO. Here we have assumed for simplicity a constant r(LOX) = 0.15, a value based upon our simulation at z$_{em} \sim${} 3.5. The rapid rise of the number of expected lines in Figure 1 is a result of the almost linearly growing redshift pathlength $\int \frac{dX}{dz}$dz. It is clear from Figure 1 that beyond redshifts of z$_{abs} \sim${} 4.2, the sheer number of interlopers will make unambigious detections of \ion{O}{6}{} features very difficult, and the search efficiency will peak below that redshift.   
\subsection{Summary of the initial feasibility analyses}
From the simulations and estimates above, we conclude that we will be able to retrieve a large enough sample of \ion{O}{6}{}  absorbers in the spectra of SDSS quasars to obtain meaningful results for a statistical analysis. Furthermore, we will be able to exclude false identifications via a variety of methods, so that we can hope to clean, to a certain degree, our sample of \ion{O}{6}{}  system candidates up to redshifts $z_{em} \sim${} 4.0.


\section{The AGN sample}\label{agn_sample}
\subsection{SDSS imaging and spectroscopy}
The Sloan Digital Sky Survey (SDSS) is an imaging and spectroscopic survey of the sky \citep[]{york2000} using a dedicated wide-field 2.5m telescope \citep[]{gunn2006, gunn1998} at Apache Point Observatory, New Mexico. Imaging is carried out in a drift-scan mode using five broad $u g r i z$ bands spanning the range from 3000 to 10,000 \AA{}  \citep[]{fukugita1996}. The 95\% completeness limit in the $r$ band is 22.2 mag (on the AB system). For more specific information on the imaging and quality control thereof see \citet[]{hogg2001, ivezic2004, smith2002}{} and \citet[]{tucker2006}. Details of the astrometry can be found in \citet[]{pier2003}. A variety of algorithms selects objects from the imaging data for spectroscopy. These targets are arranged on series of plug plates, called tiles, with radius 1.49$^{o}$ \citep[]{blanton2003}, each containing provision for a total of 640 targets and calibration stars. Optical fibers at the focal plane feed the light from holes in aluminium plates to a pair of double spectrographs. The resulting spectra range from 3800 to 9200 \AA{} with a resolution of R = $\frac{\lambda}{\Delta\lambda} \sim 1800$. The S/N of the SDSS spectra entering the public data base are required to be above 4.0 at a fibre magnitude of 19.9 in the red SDSS spectrum and a g magnitude of 20.2 in the blue \citep[]{adelmanmccarthy2006}. Specific information on SDSS data products can be found in \citet[]{stoughton2002} and \citet[]{abazajian2003, abazajian2004, abazajian2005} as well as \citet[]{adelman2007}.

\subsection{The SDSS absorption-line catalogue}
Quasars within the SDSS survey are mainly selected by colour criteria from the imaging campaign, and subsequently verified by spectroscopic follow-up studies \citep[]{richards2002}. Quasar spectra with absorption lines have been published by a number of groups : \citet[]{hall2002a, menou2001, tolea2002, reichard2003, hall2002b}, mainly focusing on broad absorption features. Several teams have developed pipelines isolating and identifying absorption lines and systems in QSO spectra \citep[]{vandenberk2000, richards2001, york2001,york2005}. Following the format of the catalogue by \citet[]{york1991}  a comparison of these pipeline-created catalogues with lists made by visual inspection reveal a completeness of above 95\%{} for such an automation. The incompleteness results mainly from blending of the \ion{Mg}{2}{} line, key to the detection of QSO absorption systems, most importantly with night sky emission.\\
The main catalogue of \citet[]{york2005}{} presents lists of significant absorption lines, equivalent widths, various line parameters and possible redshifts. Discrete absorption line systems are obtained by matching lines of different transitions of redshifts within a certain range, depending on the S/N of the spectra. Different criteria such as number and quality of components allow for a grading of the systems - all Class A systems e.g. need to contain at least four, unblended lines above a 4$\sigma${} detection limit matching in redshift, free of artefacts and blends, occuring redward of the Lyman $\alpha${} forest \citep[]{york2005}.
The QSOALS catalogue used here is based on the SDSS data release 3(DR3) \citep[]{abazajian2005}.

\subsection{Our SDSS QSO sample}
At rest-frame wavelengths of 1031.912 \AA{} and 1037.613 \AA, the \ion{O}{6}{}  doublet starts to become visible in the spectra of SDSS quasars only at redshifts of z$_{abs} \sim${} 2.7. We have thus retrieved all normalised QSO spectra in the QSOALS database beyond this lower redshift. Our sample consists of 3702 quasars with redshifts ranging from 2.70 to 5.413 \citep[]{schneider2005}. Figure 3 shows the redshift distribution of the sample. More than 60\%{} of the sources are below a redshift of z$_{em} =${} 3.5, where we expect to maximise the search efficiency for \ion{O}{6}{}  doublet signatures and the cleanliness of the \ion{O}{6}{}  candidate sample. As Figure 3 demonstrates, the majority of the sources have $i${} band magnitudes between 19.5 and 20.5. As the reliability of an absorption line detection is a function of the local signal-to-noise ratio, we have determined the average signal-to-noise ratio of all spectra within the region of interest for us. This average value is usually below the overall signal-to-noise ratio of the spectrum since the flux levels within the Lyman $\alpha${} forests are much lower than beyond it. Figure 3 shows the distribution of these average signal-to-noise ratio values with the emission redshifts of the sample sources. Especially towards higher redshifts, it is apparent that the increasing density of the forest and the general faintness of the sources lead to a severe decrease in the signal-to-noise ratio that will hinder unambigious identifications of absorption lines. However, below z$_{em} \sim${} 3.5 there are enough quasars with signal-to-noise ratio above the conservative limit we used in section 2.1 for the feasibility study : 359 spectra are of objects with z$_{em} < 3.5${} and have a signal-to-noise ratio within the forest of S/N$>$5.0.

\begin{figure}
\includegraphics[angle=270,width=\columnwidth]{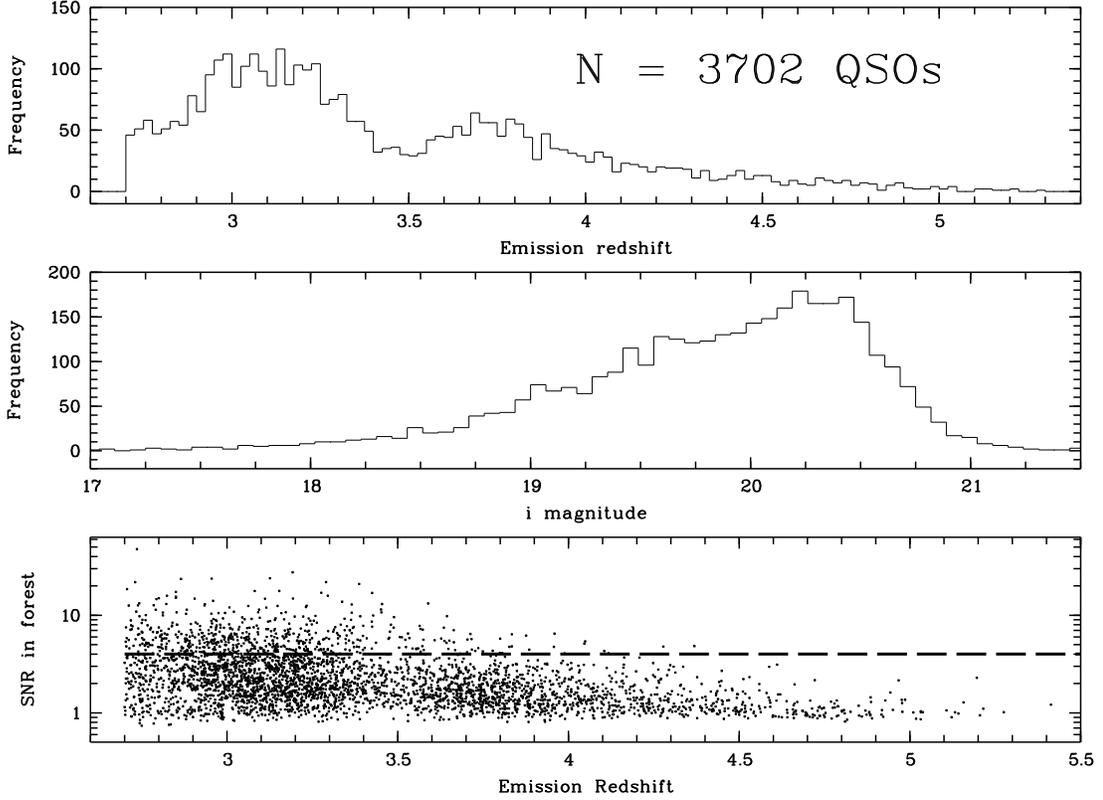}\label{z_distribution}\caption{{\bf Upper panel :} The redshift z$_{em}${} distribution of our SDSS QSOALS sample of 3702 quasars. Note that more than 60\%{} of the sources are below a redshift of z$_{em} =${} 3.5, where we expect to maximise the search efficiency and cleanliness of the \ion{O}{6}{} candidate sample. 359 of these have a S/N $>$5.0. {\bf  Middle panel :} The $i${} mag distribution (taken from \citet[]{schneider2005}) of our SDSS QSOALS sample of 3702 quasars. The majority of the sources exhibit fluxes between 19.5 $< i \textrm{ mag } <${} 20.5. {\bf Bottom panel :} The average signal-to-noise ratio per pixel in the Lyman $\alpha${} forest for each source versus the emission redshift z$_{em}$. Due to the decreased flux in the forest, this value for the signal-to-noise ratio often falls substantially below the nominal lower limit of signal-to-noise ratio=4.0 (indicated by the dashed line) for the complete spectrum to enter the SDSS data set, especially towards higher redshifts for which the sources become fainter, in general, and also the density of the forest increases substantially.}
\end{figure}


\section{Search strategy and algorithm for the SDSS quasar spectra}\label{search_strategy}
The absorption systems and line lists of the SDSS QSOALS catalogues primarily contain securely identified features, resting primarily on key absorption lines like the \ion{Mg}{2}{}  or \ion{C}{4}{} doublets. In many cases, identification of such systems is made possible by the fact that the absorption occurs redward of the Lyman $\alpha${} forest and thus in relatively clean parts of the spectra. For our search of signatures of the \ion{O}{6}{}  doublet, however, it is clear that all of the candidates must be found in the forest. Furthermore, these \ion{O}{6}{}  systems need not be accompanied by other metal absorption systems like CIV or MgII that would allow unambigious identification outside the Lyman forest (cf. \citet[]{dave1998}). Thus, we cannot expect the automated pipelines used to create the SDSS QSOALS catalogues to have picked up on the relatively weak features within the Lyman forest in which we are interested. Therefore, we had to develop our own algorithms to construct a complete list of \ion{O}{6}{}  absorbers at intervening redshifts. After completing our search with this method, we performed a cross-check with the QSOALS database to see if the putative \ion{O}{6}{} candidates also show transitions of other ions at the same redshift. The results of these checks are summarised in section 5.1.
\subsection{The search algorithm for finding \ion{O}{6}{}  doublets}
For a median Doppler broadening parameter $b_{O VI}${} = 16 km/s, obtained by Simcoe et al. (2002), it is obvious that nearly all of the \ion{O}{6}{}  lines that we expect to be present in SDSS spectra will not be resolved, as 1 \AA{} (i.e. the rough pixel size) represents about 70 km/s in velocity space at 4000 \AA, where we expect to have the best search efficiency . Nonetheless, we implemented a simple pixel by pixel comparison\footnote{One resolution element of an SDSS spectrum consists of at least 2-3 pixels, and thus even a narrow feature like a metal absorption line will get spread out over a range of pixels. Our pixel by pixel approach here, however, works very well because we select on the equivalent width {\it ratio}, which is not affected by this dispersion. Of course, we will thus only be able to detect strong features.For a more detailed comparison of the single pixel approach versus a search after rebinning to the resolution element size, see the appendix.}{} for all spectra, characterising each single pixel within the wavelength range from the blue SDSS limit of 3800 \AA{} up to the maximum wavelength allowed for the \ion{O}{6}{}  1032\AA{} component $\lambda_{max} = (z_{em}+1) \times 1031.912${} \AA{} by its normalised flux and thus optical depth. For each pixel, we derive its hypothetical absorption redshift z$_{abs}${} as if it were the \ion{O}{6}{}  1032\AA{} component. Then we identify the corresponding pixels in the spectra that would belong to the second \ion{O}{6}{}  1038\AA{} component as well as the Lyman $\alpha$,$\beta${} and, if possible, $\gamma${} lines for that same redshift. Naturally, the overlap between the redshifted wavelength range of the \ion{O}{6}{}  1032\AA{} component pixel and the actual wavelength range of the pixel containing the corresponding \ion{O}{6}{}  1038\AA{} component is not perfect. And thus, the absorption structure of the \ion{O}{6}{}  1038\AA{} transition corresponding to its 1032 \AA{} counterpart can span in principle two pixels, and hence the {\it ratio} of the absorption measured in such pairs of 1032/1038 pixels may not be a reliable quantity for identifying \ion{O}{6}{} anymore.  The estimates of the line widths above, however, demonstrate that the velocity width of a line is most likely going to be only a small fraction of the width of a single pixel - and so this scenario will occur only in a small number of cases. Furthermore, by coincidence, the wavelength separation of the pixels in SDSS spectra and wavelength ratio specific to the \ion{O}{6}{}  doublet create an overlap at the level of 0.94 for all spectra and all pixels. We did, additionally, check for potential \ion{O}{6}{}  candidates in the \ion{O}{6}{}  pixel pairs that do overlap by only 6\%, but the number of candidates retrieved in this way agreed well with just the expectations from random interlopers. We therefore estimate that the loss of valid detections due to this complication is minimal.        
\subsubsection{Criteria for identification of \ion{O}{6}{}  lines and rejection of spurious Lyman series lines}
In order to assess whether an absorption feature exhibits the correct ratio of optical depths while not being able to fully resolve the structure of the absorption lines, we need to derive a quantitative measure of that ratio from the observed normalised fluxes per pixel. Assume that a)  the complete profile of an \ion{O}{6}{}  transition fits into one resolution element, and b) that this line is the only absorption line within that element (i.e. no blending with \ion{H}{1}{}  Lyman lines). Then, it is possible to extract the equivalent width of the \ion{O}{6}{}  in that pixel from the measured normalised flux\footnote{We rely upon the flux normalisation via the continuum estimate given in the QSOALS database.} as
\begin{displaymath}
f_{pixel, 1032\AA} = 1 - \frac{EW_{\lambda 1032\AA}}{\Delta \lambda}
\end{displaymath}
where $EW_{\lambda 1032\AA}${} is the equivalent width of the line in that pixel in the observer's frame, $\Delta \lambda${} is the pixel size in wavelength units, and $f_{pixel, 1032\AA}${} is the normalised flux of the pixel assumed to contain the \ion{O}{6}{}  1032\AA{} transition. From this estimate of the equivalent width $EW_{\lambda 1032\AA}$, we can in turn derive the expected flux $f_{exp, 1038\AA}${} at the corresponding pixel :
\begin{displaymath}
f_{exp, 1038\AA} = 1 - 1/r_{EW} \times (1 - f_{pixel, 1032\AA})
\end{displaymath}
where $r_{EW} = \frac{EW_{\lambda 1032\AA}}{EW_{\lambda 1038\AA}}$ is the ratio of equivalent widths of the two transitions. This ratio depends on the strength and velocity width of the absorber. For unsaturated lines it remains fairly close to 2, and drops to unity for saturated lines. We have modelled the \ion{O}{6}{}  1032\AA{} and 1038 \AA{} transitions with Voigt profiles of various Doppler parameters $b$, and are thus able to compute :
\begin{displaymath}
r_{EW} = \frac{EW_{\lambda 1032\AA}}{EW_{\lambda 1038\AA}} = r_{EW}(EW_{\lambda 1032\AA}, b)
\end{displaymath}
as a function of the estimated equivalent width EW$_{\lambda 1032\AA}$. This introduces a model dependence into our derivation as we need to assume a value for the Doppler parameter $b$. We find, however, that even large variations of the Doppler parameter, $5.0 km/s < b < 20.0 km/s$, only affect $r_{EW}${} on the 30\%{} level. Here we take $b = 15 km/s${} as fiducial value for the Doppler parameter.\\ 
The following criteria were used to classify pixels as candidates belonging to an \ion{O}{6}{}  doublet.
\begin{enumerate}
\item Both pixels of the \ion{O}{6}{}  doublet candidate need to exhibit normalised fluxes lower than 1.0. This simply ensures that absorption features rather than pixels whose noise pushed them over the flux level of unity are selected.
\item The expected flux $f_{exp, 1038\AA}${} is allowed to deviate from the  actually measured flux $f_{pixel, 1038\AA}${} less than a certain threshold
\begin{displaymath}
\Delta f = \left|f_{exp, 1038\AA} - f_{pixel, 1038\AA}\right| \leq k_{1} \times \sigma  
\end{displaymath}
where $\sigma${} is the standard deviation of this flux difference, estimated from the signal-to-noise ratio of each pixel and the uncertainty on the line width mentioned above. The effects of varying $k_{1}${} will be described in the next section.
\item The inferred equivalent width of the stronger \ion{O}{6}{}  1032\AA{} component has to be above a minimum limit
\begin{displaymath}
EW_{\lambda 1032\AA, rest-frame} \geq k_{2} \times \Delta \lambda
\end{displaymath}
This enables us to clear our candidate lists from very weak features, i.e. spurious Lyman $\alpha${} forest lines with low column densities which we expect to be the main contaminant at high redshifts. Note that this criterion is usually automatically fulfilled for candidates in spectra with low signal-to-noise ratio passing criterion 4 mentioned below, and thus is mostly useful for cases where the signal-to-noise ratio is significantly above the 4.0 threshold for entering the sample.
\item The significance of absorption\footnote{Note that our definition for ``significance'' relies upon the following definitions : our signal is the strength of absorption in each pixel, i.e. 1.0 - f$_{pixel}${} in the normalised units. The ratio of this quantity to the noise in the same units, i.e. $(1/SNR) \times f_{pixel}$, is our metric for the significance.}{} in each pixel should be above a low threshold
\begin{displaymath}
1.0 - f_{pixel} \geq k_{3} \times \textrm{ 1/signal-to-noise ratio(pixel) } \times f_{pixel}
\end{displaymath}
While this criterion is already implicitly in place for the expected and measured flux difference and the lower limit to the equivalent width, this formulation puts a stronger limit on the number of candidates. Requiring a high significance for each pixel allows us to focus on absorption lines with a high likelihood of being real, rather than noise artifacts. In contrast to criterion 3, this cut proves most useful in spectra with low signal-to-noise ratios. Note that this criterion introduces a lower threshold for the equivalent width that is signal-to-noise ratio-dependent : at a fiducial average signal-to-noise ratio of 5.0 (8.0) and a significance level of 4.0 $\sigma${} (2.0 $\sigma$), it corresponds to a rest-frame equivalent width limit for the SDSS set-up of EW$_{rest}$ = 0.195 \AA{} (0.06 \AA).  
\item Since we expect \ion{O}{6}{}  to be accompanied by \ion{H}{1}{}  Lyman features, we require
\begin{displaymath}
f_{Lyman x} \leq f_{x} \textrm{(limit)}
\end{displaymath}
where the x represents either Lyman $\alpha, \beta${} or - if applicable - $\gamma$. Due to a lack of information regarding the metallicity and the physical state of the absorbing gas, there is, a priori, no straight forward way of calculating the \ion{H}{1}{}  absorption expected for a given value of $f_{pixel, 1032\AA}$. Therefore, we introduce these hard cuts on the flux for the associated Lyman series transitions, and resort to this simple, yet efficient method of checking for accompanying \ion{H}{1}{}  absorption.
\item We reject all candidates that show signs of being part of Lyman series absorption, i.e. we check (by eye) whether the feature in question could be a Lyman line, accompanied by higher (or lower) series members elsewhere in the spectrum with a similar velocity structure. This turns out not to be a severe cut : only 4 of altogether 1760 candidates passing all other tests had to be eliminated by this procedure. Thus, we find it unnecessary to apply a more rigorous test than this check by eye.      
\end{enumerate}
It is clear that the efficiency of detecting \ion{O}{6}{}  candidates and filtering out spurious \ion{H}{1}{}  Lyman series absorbers is a complicated function of redshift and signal-to-noise ratio of each spectrum. Thus, a more flexible approach, tailored to each object, could result in a higher overall yield. We have, nevertheless, applied the aforementioned rigorous, yet simple strategy to the complete sample regardless of redshift and signal-to-noise ratio of each source. This allows for a more straightforward comparison of the results with expectations based upon our initial feasibility study, and certainly facilitates quantitative analyses like completeness estimates. And it is certainly the conservative approach.\\
We note that possible blends of the two \ion{O}{6}{} 1032/1038 \AA{} transitions with other lines at very similar rest-wavelengths constitute an additional complication. Specifically, the presence of strong lines of \ion{C}{2} might have the 1036 \AA{} line of \ion{C}{2} overlapping with \ion{O}{6}{} 1038 \AA, and potentialy shift this blend to a lower wavelength. Also, any system that contains appreciable amounts of molecular hydrogen, H2, will also produce a blend with the \ion{O}{6}{} 1038 \AA{} transition, but leave the 1032 \AA{} component untainted\footnote{In fact, there are also two H2 components at 1032.191 and 1032.356 \AA, but these are much weaker than the 1038 \AA{} transitions.}. At the redshift range of interest for us, however, the separation between the \ion{C}{2}{} and \ion{O}{6}{} 1038 \AA{} line centres amounts to at least 6 \AA (observed frame), too much to seriously affect our search criteria. Hence, we have not implemented any correction factors due
to the presence of these features into our search algorithms. \\
Interestingly enough, none of our good candidates appear in systems that are bona-fide damped-Lyman $\alpha${} absorbers (DLAs). While Fox et al. (2007) argue that potentially all DLAs contain appreciable amounts of \ion{O}{6}{} absorption, we caution that in their sample of 35 high-resolution spectra of DLAs and sub-DLAs covering also the OVI doublet, about 60\%{} of the sightlines show blended features due to the high density of forest lines. 
\subsubsection{Effects of varying the selection criteria on the number of candidates}
Obviously, the ability to retain most real \ion{O}{6}{}  absorbers in the candidate list, while filtering out the contaminating spurious \ion{H}{1}{}  features mimicking metal lines, depends crucially on the choice of the parameters k$_{i}$ and $f_{x}$(limit) from our search criteria. We have extensively tested the effects of different choices for these parameters on the numbers and the quality of the candidates. 

\begin{figure}
\includegraphics[angle=270,width=\columnwidth]{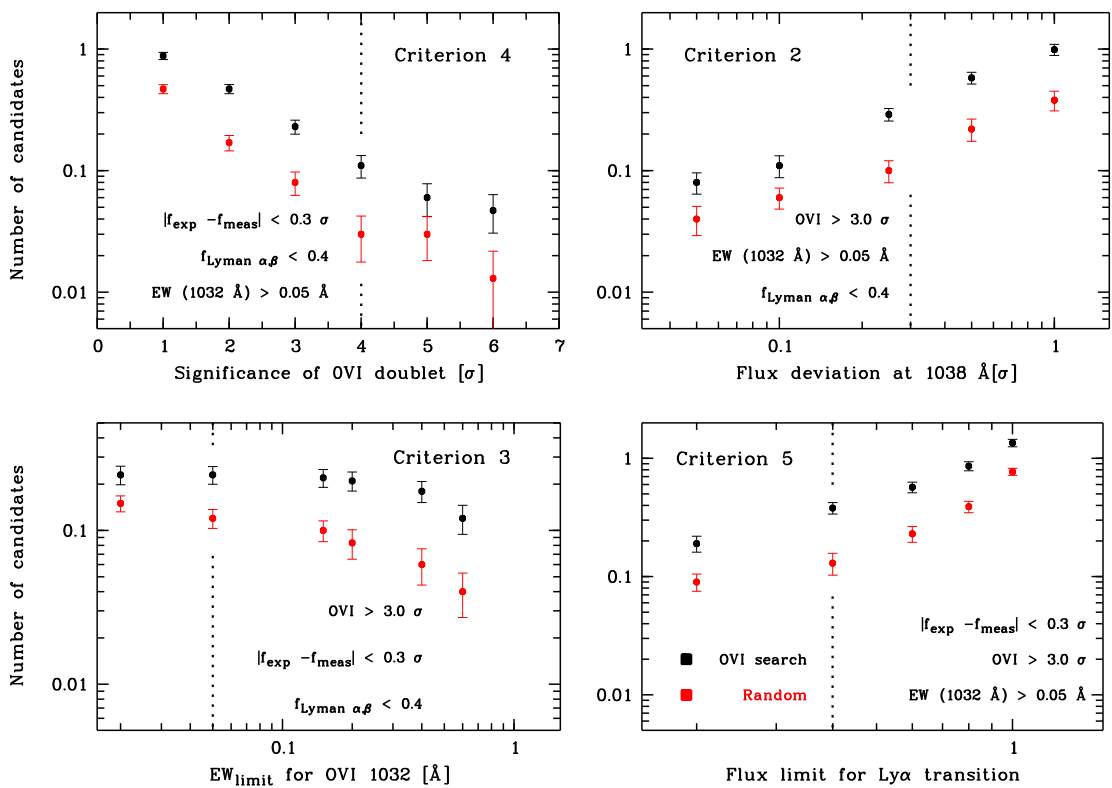}\label{cut_effects}\caption{The effects of varying cut criteria on the average number of candidates per spectrum retrieved from all 380 spectra of the sample within the redshift range $2.9 \leq z_{em} \leq 3.0$. For all panels, the data points in black result from the fully implemented search algorithm for \ion{O}{6}{} candidates, whereas data points in red are obtained by searching for candidates with random redshift separations instead of the fixed one based upon the \ion{O}{6}{} wavelength ratio, while still maintaining all the other criteria. The dotted, vertical lines indicate the choices for the cut criteria that minimise the spurious interloper contribution, while simultaneously maximising the number of real candidates kept in the sample. {\bf Criterion 4} (upper left panel) The ratio of the numbers of real to spurious candidates increases up to a significance level of $\sim 4 \sigma$. {\bf Criterion 2} (upper right panel) The average number of candidates retrieved is a strongly increasing function of the allowed flux deviation of the \ion{O}{6} 1038 \AA{} pixel from the expected value. The uncertainty in the Doppler width of the absorbers contributes significantly to the error allowance in the estimate for the expected flux. {\bf Criterion 3} (lower left panel) The lower limit on the rest-frame equivalent width of the \ion{O}{6}{} 1032 \AA{} component appears well before the turn-over to a much lower retrieval rate of real candidates ($\sim${} 0.3 \AA{} for 2.9 $\leq z_{abs} \leq${} 3.0). Criterion 4 is a stronger selector for the low signal-to-noise ratio spectra, thus we have chosen this weaker selector mainly to avoid very weak interlopers for candidates in high signal-to-noise ratio spectra. {\bf Criterion 5} (lower right panel) Requiring a low flux value for the accompanying Lyman transitions, removes effectively a large fraction of the random \ion{H}{1}{} lines. Note that applying even stricter limits will throw out good candidates in low signal-to-noise ratio spectra when noise can push a Lyman pixel over the threshold limit.}  
\end{figure}

Figure 4 demonstrates how changing one of the cut criteria while keeping all others constant affects the average number of candidates retrieved from the 380 spectra within the  redshift range $2.9 < z_{em} < 3.0$. These spectra have on average about 300 usable pixels. The data points in black indicate the average number of candidates per spectrum when applying the full search strategy, while the points in red are derived by introducing a random shift of $\delta \lambda$(random) rather than the true $\delta \lambda (z_{abs})$ for the doublet separation, but keeping the other selection criteria (accompanying Lyman transitions, line significance and equivalent widths). This procedure effectively measures the frequency of interloper incidences. Note that in all cases, we retrieve more candidates for the 'real' search than for the randomised case.  We have chosen the following parameters for filtering our dataset 
\begin{enumerate}
\item Criterion 2 $(k_{1}) :  \Delta f \leq 0.3 \times \sigma$\\While at first hand, it seems unnecessary to apply such a strict criterion, an inspection of the candidate lists produced with softer limits revealed that too many spurious candidates could pass. This is mainly due to the uncertainty in the estimated, expected EW of the \ion{O}{6}{}  1038\AA{} feature when not knowing the doppler width  of the absorbers. While obviously potentially eliminating good candidates in slightly more noisy spectra, the gain in clearing the sample from \ion{H}{1}{}  absorbers mimicking \ion{O}{6}{}  clearly overweighs.
\item Criterion 3 $(k_{2}) : EW_{\lambda 1032\AA} \geq 0.05 \times \Delta \lambda$\\The lower limit on the equivalent width of a line introduced by this cut occurs well before the turn-over to a much lower retrieval rate (at EW $\sim$ 0.3 for the case of 3.9 $<${} z$_{abs} <${} 4.0, see figure 4), but still eradicates effectively a higher number of spurious very low column-density \ion{H}{1}{} absorbers that tend to become ubiquitous at higher redshifts.  We have chosen this low limit mainly to keep the (few) cases of spectra with very high signal-to-noise (for SDSS standards, i.e. signal-to-noise ratio$_{forest} \geq${} 10.0).  Note that the lower EW limit introduced by this cut depends on the wavelength coverage of a pixel, and is thus slightly redshift dependent.
\item Criterion 4 $(k_{3}) :  1.0 - f(pixel) \geq 4.0 \times 1/$ S/N(pixel)$\times f_{pixel}$.\\The number of candidates is a strongly declining function of the detection significance, as demonstrated by figure 4. However, above a significance level of $\sim 4.0 \sigma${} the curve shows signs of flattening, before, finally, at even higher values one simply runs out of pixels above the required signal-to-noise ratio. Thus we have decided to apply the above limit, which turns out to be the strongest selector in cutting down the numbers of candidates. 
\item Criterion 5 : Flux limit for Lyman lines
 \begin{itemize}
 \item$f_{Lyman \alpha} \leq 0.4$
 \item$f_{Lyman \beta} \leq 0.6$ (if applicable)
 \item$f_{Lyman \gamma} \leq 0.8$ (if applicable)
 \end{itemize}
 Limiting ourselves only to candidates that are accompanied by potential Lyman features of such depth maximises the ratio between real \ion{O}{6}{} absorbers and  random interlopers, as can be seen from figure 4. Requiring even lower flux values throws out mainly potential candidates in low signal-to-noise ratio spectra, when noise pushes a Lyman pixel over the threshold, while softening the flux criterion leaves too many spurious interlopers in the sample.
\item In a final step, all of the candidates having passed the above criteria, were checked for possible higher (and also lower) associated Lyman series lines.  Only 4 (out of 1756) candidates had to be eliminated this way, which we take as a sign of having implemented already rigorous cuts to the sample with the criteria discussed above.   
\end{enumerate}   

\section{\ion{O}{6}{}  doublet candidate list}\label{candidate_list}

\begin{figure}

\includegraphics[angle=270,width=\columnwidth]{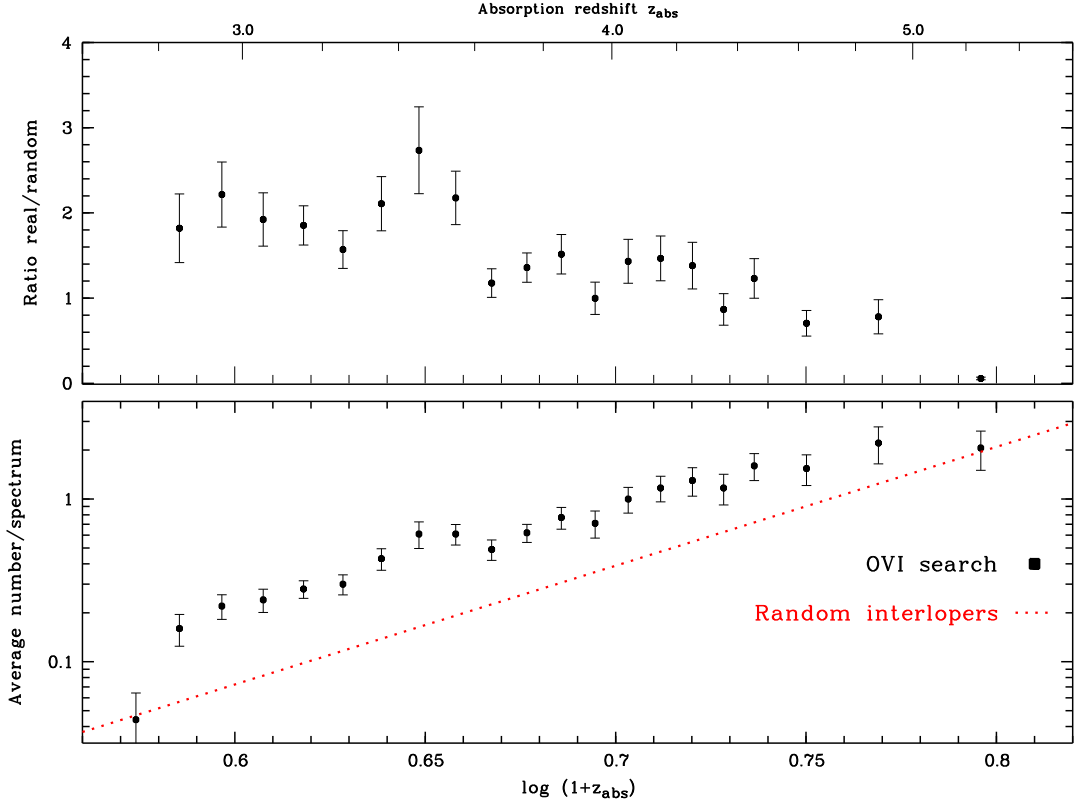}\label{results}\caption{The results of the 'real' and 'random' search algorithms applied to the 3702 QSO sample. This graph can be directly compared to Figure 1 where the expected values based upon theoretical estimates for the Lyman $\alpha${} forest density and the metallicity of the IGM are presented. {\bf Lower panel:} For each redshift bin, the black data points represent the average number of candidates retrieved per spectrum. This number includes both real candidates and spurious interlopers. The red dotted line is a linear fit to the number of such random interlopers derived by averaging over 15 runs of the search algorithm when applying a random redshift offset to the location of the \ion{O}{6}{} doublet line (properly at 1038 \AA). The resulting redshift evolution of these interlopers agrees very well with the estimates from the feasibility study in paragraph 1. For details see text. {\bf Upper panel :}{} The average approximate ratio of 'real' to 'random' candidates per redshift interval. The number of 'real' candidates is assumed to be the difference between the black data points and the estimate for the interloper frequency (red dotted line in lower panel). Note the decreasing efficiency to produce a clean sample with increasing redshift. As expected from figure 1, the ``window of opportunity'' lies roughly between 0.55 $< log (1 + z_{abs}) <$0.7 (= 2.7 $< z_{abs} <$ 4.0), when the ratio real/random rises above unity.}
\end{figure}

Figure \ref{results}{} shows the results of applying the search algorithm with the cut criteria mentioned in section 4 to the full dataset of 3702 spectra. For each redshift bin, the black data points in the lower panel represent the average number of candidates retrieved per spectrum. This number includes both real candidates and spurious interlopers. The red dotted line is a linear fit to the number of such random interlopers derived by averaging over 15 runs of the search algorithm when applying a random redshift offset to the location of the \ion{O}{6}{} doublet line (properly at 1038 \AA). The redshift evolution of these (n$\propto (1+z_{abs})^{7.3}$) agrees very well with the expectations from the initial feasibility study (cf. equation 7 and fig. 1). Note that the number of 'real candidates' plus 'spurious interlopers' for the full search is always higher than in the random search process, lending further credibility to the robustness of our search strategy. The average approximate ratio of 'real' to 'random' candidates per redshift interval is presented in the upper panel of fig. \ref{results}. The number of 'real' candidates is assumed to be the difference between the black data points and the estimate for the interloper frequency (red dotted line in lower panel of that figure). As expected, the ability to produce a clean sample diminishes rapidly beyond redshifts of $z_{abs} \sim${} 4.0.\\

\begin{figure}
\includegraphics[angle=270,width=\columnwidth]{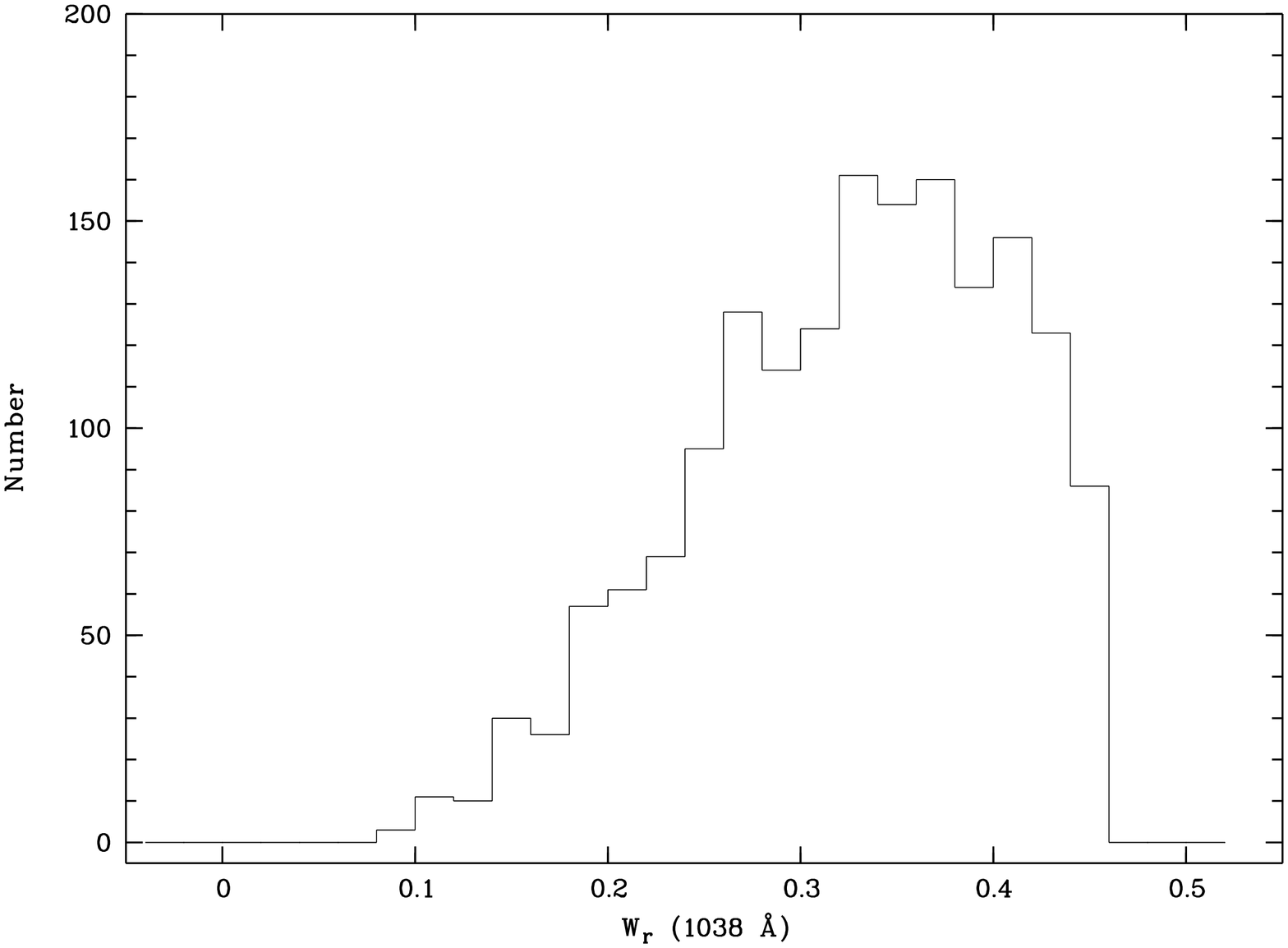}\label{EW}\caption{The rest-frame equivalent width distribution of the 1038 \AA{} component for the 1756 candidates found by our search. Note that we have 'measured' the equivalent widths in the normalised spectra by simply only including the pixel to the left and right of the candidate pixel into the calculation, and have made no attempt to take the often complicated structure of the forest around the central pixel of the candidate into account. Given that one resolution element should in almost all cases encompass the narrow OVI lines (b parameters of the order of 16 km/s) this method overestimates the width, but does provide secure upper limits.}   
\end{figure}

Altogether, we retrieve 1756 candidates\footnote{We note that in 232 cases two pixels exactly adjacent to each other were selected. Hence, the total number listed in table 2 is 1756 + 232/2 = 1866. These 116 absorber candidates are also picked up by the search after rebinning. Such cases are exactly what is expected when the noise is low enough to allow a line that will be instrumentally broadened over 2-3 pixels to retain its shape.For details see the appendix.}   in 855 different AGN spectra (out of a total of 3702) passing all of the above criteria. These candidates still exhibit a wide range of properties. We examined each case manually, and grouped them into three categories depending on the quality and likely interpretation. 
\begin{itemize}
\item {\bf High Quality Sample :}\\The highest quality, most unambigious candidates for \ion{O}{6}{} absorbers. All lines are strong and the OVI doublet plus Ly $\alpha${} and Ly $\beta${} transitions are not obviously blended. The putative \ion{H}{1}{}  absorption is both stronger than the \ion{O}{6}{} feature and exhibits the expected line ratios for Lyman $\alpha${} to Lyman $\beta${}.\footnote{We allow the Lyman $\gamma${} line to be blended or even non-existent due to its inaccessibility in the spectra for absorbers of the lowest redshift. Maintaining a strict non-blending criterion for Ly $\gamma$, which is much weaker than Ly $\alpha${} and $\beta$, would have left us with too few good candidates.} 145 candidates could be placed in this group. In 76 cases, the absorption redshift is so close to the QSO's emission redshift that these absorbers are tentatively classifiable as intrinsic ($\Delta v = c \times \frac{z_{em} - z_{abs}}{1.0+z_{abs}} \leq 5000$ km/s)\footnote{Note that the velocity separation alone does not guarantee the systems to be intervening. See \citet[]{richards1999}{} and \citet[]{misawa2007}{} for cases where high velocity absorber can be regarded as intrinsic to the AGN}. The remaining group of 69 good candidates will be the prime sample for possible follow-up studies. 
\item {\bf Medium Quality Sample :}\\These candidates generally exhibit a secure identification of many of the features in question, but also at least one characteristic that leaves some doubt, e.g. a blending of one line, weak \ion{H}{1}{}  features in comparison to the \ion{O}{6}{} or excessively broad \ion{O}{6}{} lines. 586 (30\%) candidates fall into this category.
\item {\bf Low Quality Sample :}\\In this class of the lowest quality candidates, we encounter mostly examples with very low overall signal-to-noise ratio within the forest or features that are probably strong Lyman $\alpha${} absorbers at redshifts low enough such that we cannot see Lyman $\beta${} to apply our exclusion criterion. 1029 of the 1756 candidates ($\sim 60\%$) fall into this category, which is considerably higher than the fraction we expected to be spurious interlopers from our initial feasibility studies, and thus leads us to the conclusion that even in this group there are real \ion{O}{6}{} absorbers.
\end{itemize}

Figures 7, 8 and 9 highlight examples of each of the categories.    

\begin{figure}
\includegraphics[angle=270,width=\columnwidth]{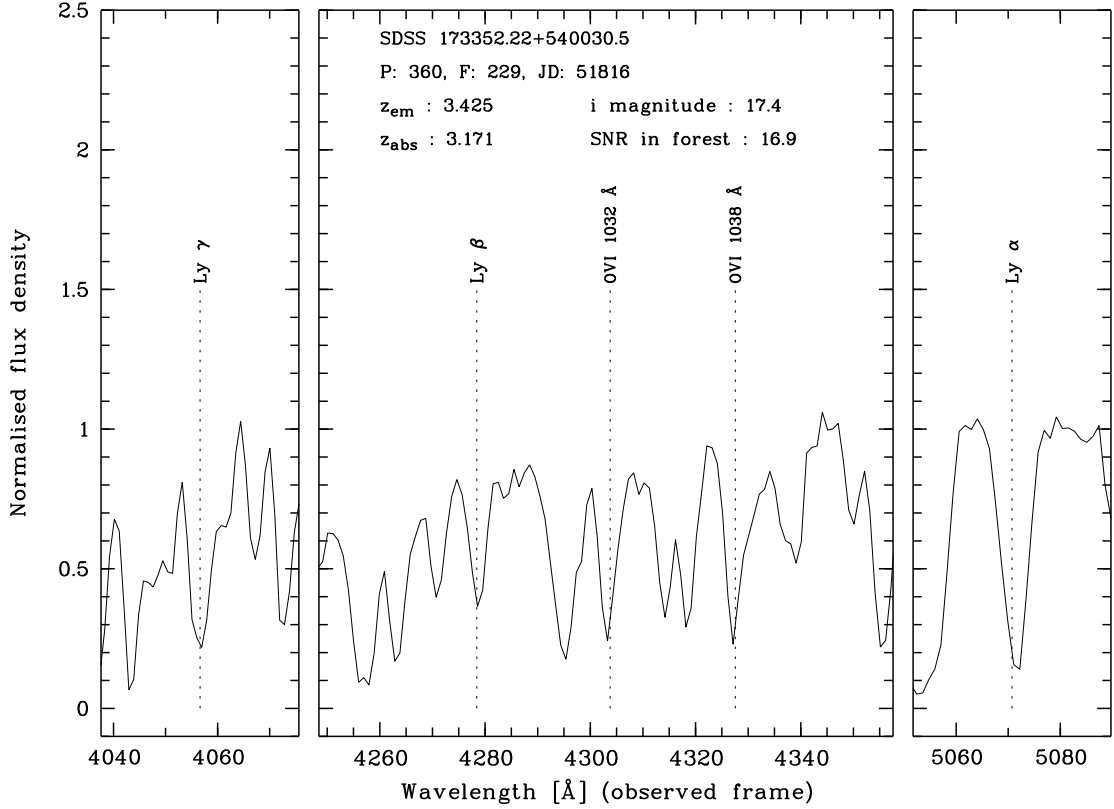}\label{figure1}\caption{Example of a high quality \ion{O}{6} absorber candidate.  All lines are strong and not obviously blended (except for the Lyman $\gamma${} feature, which we require to be present, but allow for it to be blended). The putative \ion{H}{1}{}  absorption is both stronger than the \ion{O}{6}{} feature and exhibits the expected line ratios for the Lyman series ($\alpha${} to $\beta$). The position of the \ion{O}{6}{} doublet and the associated Lyman $\alpha ,\beta , \gamma${} lines are indicated by the dotted lines. The plate, fiber and Julian date identification for the SDSS spectrum, as well as the emission and absorption redshifts together with the i magnitude and the average signal-to-noise ratio within the Lyman forest are given at the top of the graph. Note that the local S/N, used to compute the significance of a feature, may be different from the average value given here, and is always taken from the SDSS pipeline error array estimate.}
\end{figure}

\begin{figure}
\includegraphics[angle=270,width=\columnwidth]{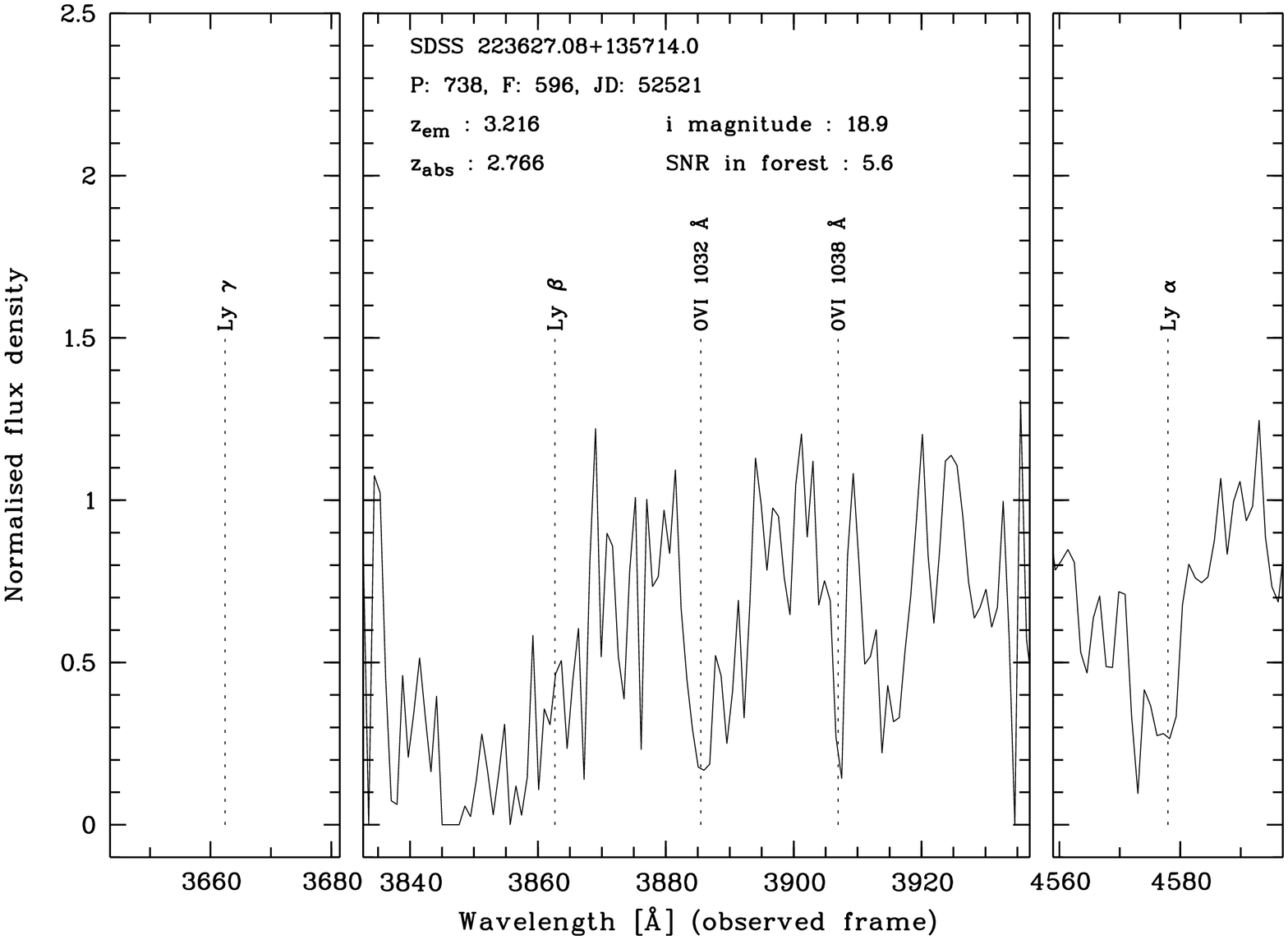}\label{figure2}\caption{Example of an intermediate quality \ion{O}{6} absorber candidate.  The position of the \ion{O}{6}{} doublet and the associated Lyman $\alpha ,\beta${} lines are indicated by the dotted lines. The plate, fiber and Julian date identification as well as the emission and absorption redshifts together with the i magnitude and the general signal-to-noise ratio within the Lyman forest are given at the top of the graph. These candidates generally exhibit a secure identification of many of the features in question, but also at least one characteristic that leaves some doubt, e.g. a blending of one line (as in the case here for the \ion{O}{6}{}  1032\AA{} component), weak \ion{H}{1}{}  features in comparison to the \ion{O}{6}{} or excessively broad \ion{O}{6}{} lines. Note that the example here also highlights a case where the absorber redshift is too low to have the Lyman $\gamma${} transition in the spectrum.}
\end{figure}

\begin{figure}
\includegraphics[angle=270,width=\columnwidth]{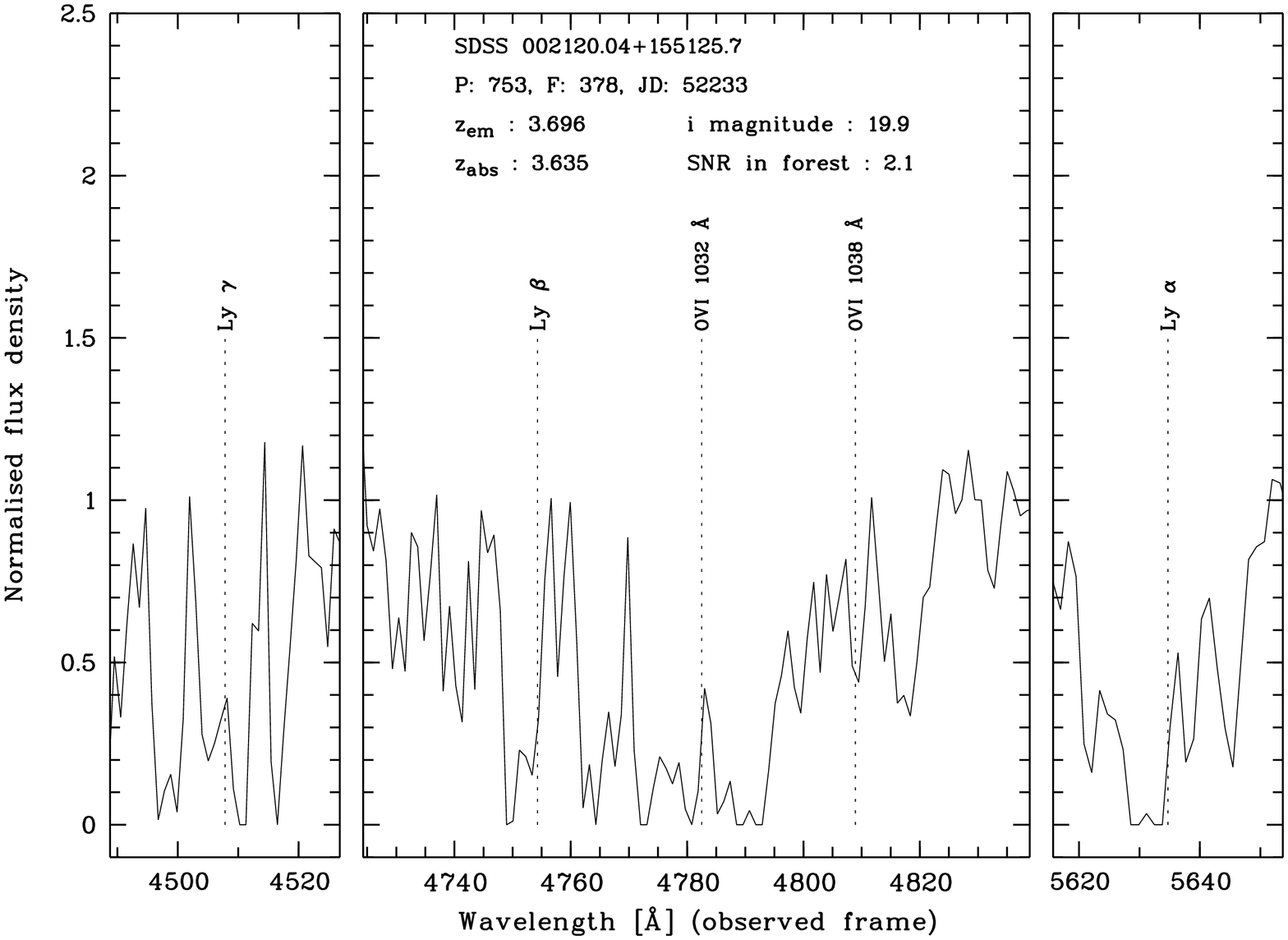}\label{figure3}\caption{Example of a poor quality \ion{O}{6} absorber candidate.  The position of the \ion{O}{6}{} doublet and the associated Lyman $\alpha ,\beta${} lines are indicated by the dotted lines. The plate, fiber and Julian date identification as well as the emission and absorption redshifts together with the i magnitude and the general signal-to-noise ratio within the Lyman forest are given at the top of the graph. These candidates generally allow only for very insecure identifications of many of the features in question, mostly because of low overall signal-to-noise ratio in the forest (as in the case presented here), but also blending and even a few cases where a strong Lyman $\alpha${} system could not be identified as such by our method due to the lack of accompanying higher order Lyman series transitions because of its low redshift. About 60\%{} of the candidates fall into this category.}
\end{figure}

\subsection{Cross-checking the candidate list with the QSOALS database}
Having constructed the candidate list of \ion{O}{6}{} absorbers with the search algorithm described above, we performed a cross-check with the QSOALS database in order to determine whether there are additional ions seen at the same redshift. Here, we have focused first on the 69 candidates for intervening absorbers of the highest quality, but will extend the cross-check fully in a forthcoming  paper. We have checked for other metal lines accompanying the \ion{O}{6}{} candidates within a range of $\pm 800$ km/s.  For 10 of these candidates the QSOALS database lists secure identifications of absorber systems at redshifts very close to value determined from the \ion{O}{6}{} doublet. In all of these cases, the velocity difference between the \ion{O}{6}{} absorber determined by our search algorithm and the listed QSOALS redshift for the absorber system is less than 300 km/s. These absorbers show transitions in a variety of other ions, most prominently \ion{C}{4}{} 1548/1550 \AA, \ion{C}{2} 1335 \AA, \ion{Si}{4} 1394/1402 \AA, \ion{Al}{2} 1671 \AA, and, in 2 cases of low enough redshift not to push the lines beyond the SDSS spectral window, even \ion{Al}{3}{} 1855/1863 \AA. Not only does the existence of the other transitions in these absorbers greatly increase the likelihood for the \ion{O}{6}{} absorbers to be real, but it will also enable us to study the physical state of the absorbing medium via the full, established  arsenal of astrophysical diagnostics. The in-depth analysis of these absorbers will be the central theme of forthcoming papers.\\
While the remaining 59 candidates have no absorption systems listed in the QSOALS database, there are cases where the database indicates the detection of an absorption feature at the location expected from our \ion{O}{6} redshift, but does not assign an identification, probably due to a failure to securely identify the transition. We have checked for the presence of the following ions : \ion{C}{4}{} 1548/1550 \AA, \ion{C}{2} 1335 \AA, \ion{Si}{4} 1394/1402 \AA, \ion{Al}{2} 1671 \AA, and \ion{Al}{3}{} 1855/1863 \AA. These are the primary transitions that are expected to be strong enough and, due to their rest wavelengths, can occur beyond the Lyman forest in the part of the spectrum where the detection is thus much easier. For 36 out of the 59 candidates, we do not find any securely detected absorption feature in the QSOALS catalogues. In 11 cases, there is one of the transitions listed, and in 8 cases 2 absorption features are present. For another two absorbers, we found 3 features each, and in the two best cases, even 4 of the 8 transitions in question could be retrieved. Given the low likelihood that these features are noise artifacts, especially for the cases where they are redward of the Lyman forest, the additional presence of up to 4 absorption lines lends further credibility to the reality of the \ion{O}{6}{} candidate.\\
We note that the non-detection of a transition in the QSOALS catalogue cannot be equated with the non-existence of the feature in the spectrum. In order to enter the QSOALS catalogue, a feature needs to pass certain threshold criteria, set by the automatic program detailed in \citet[]{york2005}. Specifically, the requirement of a four-sigma detection of a feature results in an equivalent width sensitivity which depends on the signal-to-noise ratio, but is at least of order 50-200 m\AA{}  in the absorber restframe \citep[]{york2005}. We are going to manually recheck each spectrum separately at the wavelengths where we expect potential absorption features, and derive upper limits for lines not detected. These can also help with the interpretation of the physical state of the absorbing gas.\\
A more difficult task is the assessment of the completeness level of our sample. We have noticed that, in some cases, the QSOALS database indicates the presence of absorbing systems with \ion{O}{6}{} which is not being picked up by our search algorithm. We have examined closely the results for both search algorithms for the 134 QSO sightlines with 3.2 $\leq z_{em} \leq${} 3.23. We find that we retrieve 12 out of 37 grade A, B and C absorption systems (containing 4, 3 or 2 secure line identifications at the same redshift, respectively)\footnote{For the exact definitions of these grades see \citet[]{lundgren2009}.} with \ion{O}{6}{} doublet identification in the QSOALS, whereas none of the 5 grade D systems with \ion{O}{6} is detected by our algorithm. On the other hand, we obtain 24 \ion{O}{6}{} candidates in that subsample that do not appear in the QSOALS database, or only show one of the two doublet components. \\
What is the reason for the apparently low ($\sim$ 30 \%) success rate for our algorithm compared to the QSOALS search program ? In most of the cases where we do not detect a system, one or even both of the \ion{O}{6}{} doublet transitions are blended, and in some cases there was a lack of accompanying Lyman $\beta${} absorption. Furthermore, in a significant fraction of the QSOALS candidates, there is a velocity difference between various metal components greater than the SDSS pixel separation, as pointed out by York et al. (2006), which leads to an exclusion of the candidate in our approach. Thus, it is clear that the rigidity of our search criteria does lead to a certain fraction of lost real candidates. In order to meaningfully constrain the properties of the IGM at high redshift, the estimate of this lost fraction is absolutely crucial.  We postpone a full Monte-Carlo analysis of the search efficiency to the next paper in the series that will deal with this issue in depth for the absorbers with the highest likelihood of being real.\\
The results of this crosscheck with the QSOALS database are summarised in Table 3.

\begin{table}\label{summary_of_surveys}\caption{Overview of searches for intervening \ion{O}{6}{} absorbers at high redshift.}
\begin{center}
\begin{tabular}{cccc}
Redshift range & Number of  & Number of    & Reference \\
                            & QSO sightlines & \ion{O}{6}{} absorbers & \\
\hline
\hline
z$_{abs} \sim$ 0.9          & 11                               & 6 systems                                                                 & Burles \&{} Tytler (1996) \\
1.46 $< z_{abs} <$ 1.81 & 1                                 & 6 systems                                                & Reimers et al. (2006) \\
                                              &                                     & 8 additional candidates                         & \\
z$_{abs} \sim$ 2.5           & 7                                 & $\sim$ 50 systems                                                & Simcoe et al. (2004) \\
2.0 $< z_{em} <$ 2.5     & 10                               & 136 (component) candidates                                          & Bergeron \& \\
                                            &                                      & in 51 systems                                           & Herbert-Fort(2005) \\
2.0 $< z_{abs} <$ 2.36             & 2                                  & 20 (component) candidates                                       & Carswell et al. (2002) \\
                                                &                                     & in 12 systems                                         & \\
2.1 $< z_{abs} <$ 3.1                                           & 35 DLA systems  & 9 (system) candidates               & Fox et al. (2007) \\
z$_{abs}$ = 2.3                 & 2 lensed QSO pairs & 10 components                                                              & Lopez et al. (2007) \\
\hline
\hline
2.7 $< z_{abs} <$ 4.2     & 3702                         &  1756 candidates (systems)                                                           & This work. \\
                                               &                                   &  145 Category 1 candidates,                &                                     \\
                                              &                                    &  69  of these intervening                       &                                        \\
\hline

\end{tabular}
\end{center}
\end{table}


\section{Summary and conclusions}\label{summary}
We have systematically searched for signatures of metal lines in quasar spectra of the Sloan Digital Sky Survey (SDSS), focusing on finding intervening absorbers via detection of their \ion{O}{6}{}  doublet. In this paper, we have presented our search algorithm, and criteria for distinguishing candidates from spurious Lyman alpha forest lines. In addition, we compare our findings with simulations of the Lyman alpha forest in order to estimate the detectability of \ion{O}{6}{}  doublets over various redshift intervals. We have obtained a sample of 1756 \ion{O}{6}{}  doublet candidates in 855 AGN spectra (out of 3702 objects with redshifts in the accessible range for \ion{O}{6}{}  detection). This sample is further subdivided into 3 groups according to the potential for follow-up of these candidates : 145 of the candidates with the cleanest signatures for \ion{O}{6}{}  doublets with high signal-to-noise and high resolution are promising candidates for higher resolution spectroscopy in order to better constrain the physical state of the absorbers. Seventy-six of these, however, could be intrinsic absorbers as their redshift and thus velocity separation from the QSO is less than 5000 km/s. This leaves us with 69 strong candidates of intervening \ion{O}{6}{} absorbers at redshifts beyond $z_{abs} =$ 2.7.\\

The efficiency of detecting \ion{O}{6}{}  candidates and filtering out spurious \ion{H}{1}{} Lyman series absorbers is a complicated function of redshift and signal-to-noise ratio of each spectrum, and thus more flexible search criteria, tailored to each object, could result in a higher overall yield than achieved here. We decided, however, to apply rigorous, yet simple cuts to the complete sample regardless of redshift. This will allow for a more straightforward comparison of the results with expectations based upon our initial feasibility study, and certainly facilitates quantitative analyses like completeness estimates, which we will undertake in a forthcoming paper.\\
To summarise, our ``blind'' pixel-by-pixel search for \ion{O}{6}{} absorbers in the over 3700 SDSS spectra of redshifts beyond z$_{em}${} = 2.7 complements other studies that have targeted possible \ion{O}{6}{} features by their association to already known, strong absorbers such as damped Lyman $\alpha${}, Lyman limit systems and even metal-line absorbers \citep[]{simcoe2002, simcoe2004}. Traditionally, these studies have focused on very high signal-to-noise ratio and resolution, at the price of thus being limited to relatively few individual sources due to the expensive nature of the observations. Our sample, residing at z$_{abs} \geq 2.7$, therefore greatly increases the redshift range of known \ion{O}{6}{} absorbers at high redshifts. It will allow us to test expectations  based upon photoionisation models like the ones presented in \citet[]{dave1998}, thereby possibly constraining the physical conditions of the absorbers and the metagalactic UV/X-ray background at high-redshifts.\\

We would like to thank the anonymous referee for valuable comments that have improved the quality  of the manuscript dramatically. \\

Funding for the SDSS and SDSS-II has been provided by the Alfred P. Sloan Foundation, the Participating Institutions, the National Science Foundation, the U.S. Department of Energy, the National Aeronautics and Space Administration, the Japanese Monbukagakusho, the Max Planck Society, and the Higher Education Funding Council for England. The SDSS Web Site is http://www.sdss.org/.

The SDSS is managed by the Astrophysical Research Consortium for the Participating Institutions. The Participating Institutions are the American Museum of Natural History, Astrophysical Institute Potsdam, University of Basel, University of Cambridge, Case Western Reserve University, University of Chicago, Drexel University, Fermilab, the Institute for Advanced Study, the Japan Participation Group, Johns Hopkins University, the Joint Institute for Nuclear Astrophysics, the Kavli Institute for Particle Astrophysics and Cosmology, the Korean Scientist Group, the Chinese Academy of Sciences (LAMOST), Los Alamos National Laboratory, the Max-Planck-Institute for Astronomy (MPIA), the Max-Planck-Institute for Astrophysics (MPA), New Mexico State University, Ohio State University, University of Pittsburgh, University of Portsmouth, Princeton University, the United States Naval Observatory, and the University of Washington. 

\bibliographystyle{aj}

\section{Appendix A : Candidate list of OVI Absorbers and Details for best candidates}
The following table is an excerpt of the list of all \ion{O}{6}{} candidates retrieved from the 3702 SDSS QSO spectra in our survey. Listed are the unique identifiers of the SDSS object (plate, fiber and Julian Date of Observation), as well as the emission redshift of the QSO from the SDSS database. The overall quality flag for each absorber candidate at the given absorber redshift, z$_{abs}$, has the following meaning (for details see text) : 1 = excellent  2 = mediocre 3 = extremely poor. The signal-to-noise rate in the forest is the average S/N measured for the wavelength area of interest for the \ion{O}{6}{} search, and often substantially less the nominal limit of S/N (total) $>$ 4.0 for an object to enter the SDSS database. The velocity difference between the absorber and the QSO is calculated from the given emission line redshift estimate, z$_{em}$. Absorbers at velocity differences less than 5000 km/s (i.e. $\beta <  0.0167$) are classified here as 'intrinsic' or 'associated', while all other absorbers are tentatively 'intervening'. 
The table is sorted by increasing absorber redshift.\\
The full table is available in the electronic version.
 
\begin{table}\label{candidate_table}\caption{List of \ion{O}{6}{} absorber candidates in all 3702 SDSS QSO spectra of the survey. The quality flag has the following meaning meaning : 1 = excellent  2 = mediocre 3 = extremely poor. The complete table can be found in the online material.}
\begin{center}
\begin{tabular}{ccccccccc}
Sequence & Plate  & Fiber    & JD & z$_{abs}$ & z$_{em}$ & Quality flag & S/N (forest) & $\beta = \Delta_v / c$  \\
\hline
\hline
 1     &   1284   &      140   &    52736  &   2.71413 &  2.93300 &  3 &  3.1 &  0.0589 \\
2      &   935    &     592    &   52643   &   2.71670 &  3.19500 &  3 &  6.5 &  0.1286 \\
3      &   935    &     592    &   52643   &   2.71755 &  3.19500 &  3 &  6.5 &  0.1284 \\
4      &   659    &     181    &   52199   &   2.71927 &  2.94200 &  3 &  4.9 &  0.0598 \\
5      &   611    &     221    &   52055   &   2.72184 &  2.74100 &  3 &  3.4 &  0.0051  \\
6      &   659    &     181    &   52199   &   2.72269 &  2.94200 &  3 &  4.9 &  0.0589 \\
7      &   961    &     450    &   52615   &   2.73127 &  3.03400 &  2 &  5.1 &  0.0811 \\
8      &  1264    &      74    &   52707   &   2.73299 &  2.97400 &  3 &  2.8 &  0.0645 \\
9      &  291     &     612    &   51928   &   2.73557 &  2.80900 &  3 &  6.4 &  0.0196 \\
10     &   611    &     221    &   52055   &   2.73730 &  2.74100 &  2 &  3.4 &  0.0009 \\
\dots  &  \dots   &     \dots  &   \dots   &   \dots   &  \dots   &  \dots & \dots & \dots \\

\end{tabular}
\end{center}
\end{table}

\pagebreak

Table 3 contains additional information on the intervening \ion{O}{6}{} candidates with the highest likelihood of being real (category 1), and a comparison with the results of the QSOALS database. Listed are the unique identifiers of the SDSS object (plate, fiber and Julian Date of Observation), as well as the emission redshift of the QSO from the SDSS database. The signal-to-noise rate in the forest is the average S/N measured for the wavelength area of interest for the \ion{O}{6}{} search, and often substantially less the nominal limit of S/N (total) $>$ 4.0 for an object to enter the SDSS database. We have also included the position of other ionic transitions at the same redshift as the \ion{O}{6}{} doublet detected in the QSOALS database.
The first ten entries in the table are the candidates where the QSOALS database lists a secure system. The following 59 sightlines have no securely detected system at the redshift of our candidates, however, in certain cases specific single transitions are detected. \\
The full table is available in the electronic version.
 
\begin{table}\label{intervening_category1}\caption{Details for category 1 \ion{O}{6}{} absorber candidates.}
\begin{center}
\begin{tabular}{ccccccccccccccc}
Plate  & Fiber    & JD & z$_{em}$ & z$_{abs}$ & S/N (forest) & $\lambda$ CIV 1548 \AA & $\lambda$ CIV 1550 \AA & $\lambda$ CII 1335 \AA & $\lambda$ SiIV 1394 \AA & $\lambda$ SiIV 1402 \AA & $\lambda$ Al II 1671 \AA & $\lambda$ Al III 1855 \AA & $\lambda$ Al III 1863 \AA & z (QSOALS)  \\
\hline
\hline
 819 & 530 & 52409  & 3.0310  & 2.9405  & 2.1  & 6102.3  & 6114.2  & 5260.0 & 5493.7 & 5529.0 & 6587.0  &   \dots & \dots &   2.9417 \\
 971 & 508 &  52644 &  3.0568 &  2.9779 &  6.0 & 6160.1 & 6170.2 &  5308.6 & 5545.9 & \dots &  6647.7 & \dots &  \dots &  2.9791 \\ 
 830 & 431 & 52293  & 3.3880 &  3.3127 & 3.3 & 6678.4  &6689.7 & \dots  & 6012.5 & 6052.6 & 7205.3  & \dots & \dots & 3.3143 \\ 
 302 & 438 & 51688 & 3.5500 & 3.4305 & 2.3 & 6866.8 & 6877.8 & \dots &  6181.4  & 6221.3  & 7409.0 & \dots & \dots & 3.4356 \\
1165 & 391 & 52703  & 3.5790  & 3.4643  & 1.9  & 6913.3 & \dots &  5957.5  & 6223.4 &  \dots & 7459.3 & \dots & \dots & 3.4657 \\
\dots  &  \dots   &     \dots  &   \dots   &   \dots   &  \dots   &  \dots & \dots & \dots \\

\end{tabular}
\end{center}
\end{table}

\section{Appendix B : Single Pixel Search vs. Smoothing to the SDSS Resolution}

We have performed the search for OVI absorbers based upon the information in the single pixels of the original SDSS spectra. The SDSS detector is built in such a way that spectra are slightly oversampled compared to the resolution element achievable with the nominal R=1800 resolution. Hence, the question arises : Could we have done better by smoothing the spectra according to the oversampling before searching, hereby gaining potentially significantly in S/N ? The answer to this question depends to a degree on the goals that are desired for the sample of candidates. In this Appendix, we compare the pros and cons of the two possible approaches, and explain why we have chosen to use the Single Pixel Search Method.\\
Let us begin by clearly stating our criterion for a successful search algorithm. We want to obtain a sample of good quality OVI absorber candidates. With such a sample we want to be able to perform two different actions : a. the selection of follow-up candidates with higher resolution and S/N. Note that we have set out to construct a sample at higher redshift than has ever been tried before, and b.  derive basic statistical quantities such as dn/dz and density / metallicity estimates for this sample. Hence we need to try both to maximise the number of absorbers found and the cleanliness of the sample. It is immediately obvious that increasing the S/N of the basic search bin (pixel vs. resolution element) could indeed be very helpful to increase the number of absorbers we can hope to find, given the often very low quality of the spectra with S/N per pixel of the order of a few. On the other hand, smoothing over a larger bin size increases the probability of HI interlopers falling within the bin distorting the flux measurement in a good candidate, or creating new bad candidates by mimicking an OVI doublet. Therefore, it is a priori not clear which method is favourable.\\
In order to decide upon a quantitative argument, we have indeed smoothed all 3702 spectra to the nominal resolution of the SDSS spectrograph by running a Gaussian filter with the appropriate FWHM over each spectrum. The oversampling factor is about 2.5, and hence we gain roughly a factor of $\sqrt 2.5 \sim 1.6${} in S/N for each pixel. Then we re-ran just the same search algorithm over these new spectra. The results compared to the original search can be summarised as follows :
\begin{itemize}
\item In the single pixel search, we retrieve 1756 'unique' candidates for absorbers, i.e. correcting for finding double or multiple candidates within one or two pixels of each other in the same spectrum. The search in the smoothed spectra yields 5018 such candidates, when applying the same search criteria.
\item There is some overlap :  543 candidates appear in both lists, i.e. a fraction of 31\%{} of the Single-Pixel candidates are found again in the search after smoothing.  
\item The ratio of interlopers to 'all' candidates (real and interlopers), as estimated by the random search, rises to a much higher value for the search after smoothing than for the Single-Pixel search : while we estimate about 780 of the 1756 systems found via the latter method to be random interlopers (44\%), the same criteria for the results obtained with the former method yield a fraction of almost 85\%(4230 out of 5018), as detailed below.   
\end{itemize}
It is this latter point that clearly favours the implementation of the one-pixel search strategy, given the goal of a candidate list that is as clean as possible under the given noise and resolution constraints of SDSS. Note that the number of detected candidates and random interloper estimates depends strongly on the cut criteria introduced in section 4.1.1. The specific values we have adopted were chosen to achieve an optimal result for the single-pixel method. Is it possible to change those parameters for the new search in order to achieve a lower fraction of interlopers ? We have tested this by varying the cut values over a wide range of the parameter space, just like we did for the original one-pixel search as detailed in section 4.1.2., but could not find any combination of them that had the interloper fraction drop below 75\%. This is highlighted in Figure 10, that can directly be compared to Figure 4 : it is evident that the number of detections rises by a factor of about 5-8 for all cut criteria, but the fraction of random interlopers increases even more dramatically all across the board (red lines in the figure). This result is somewhat surprising given that the initial feasibility analysis showed that by increasing the S/N the ratio of real-to-random candidates should also increase. A potential complication may be the clustering of (stronger) Lyman forest features, that could increase the number of interlopers.  
\begin{figure}
\includegraphics[angle=270,width=\columnwidth]{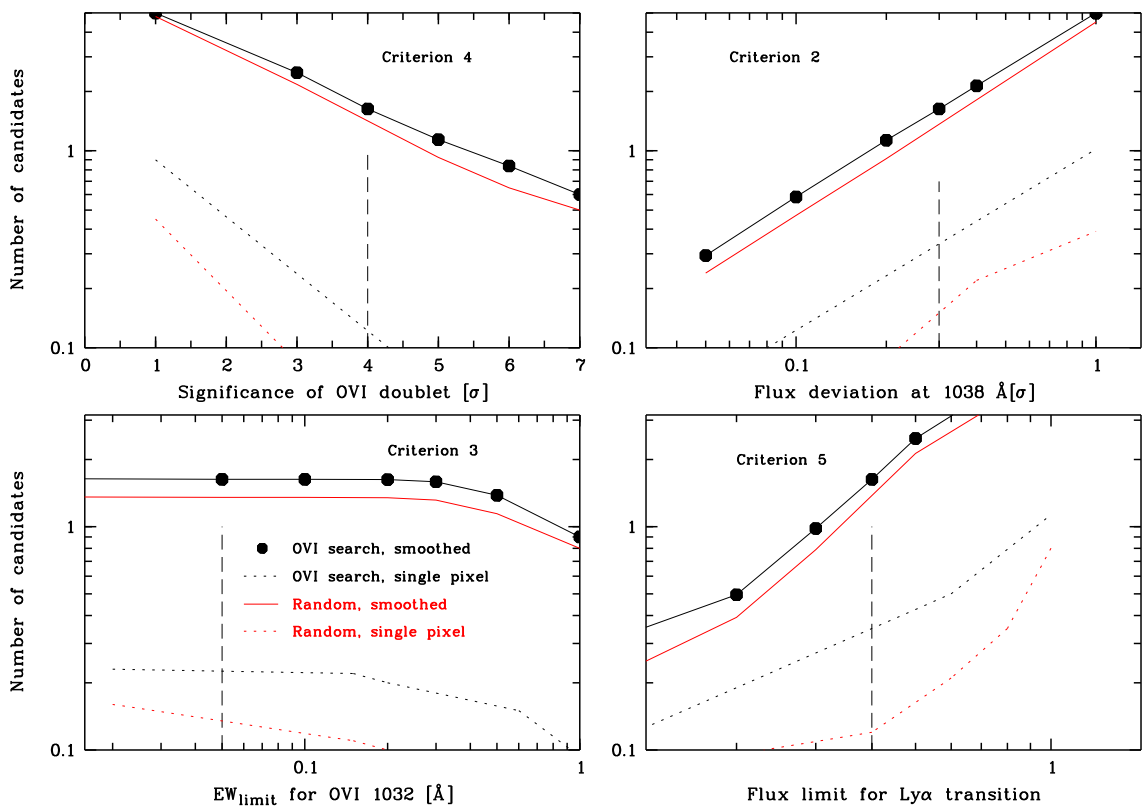}\label{result_new_search}\caption{Results of the search for OVI candidates in the spectra that have been smoothed by the proper SDSS oversampling. The number of candidates found for each combination of the cut criteria can be directly compared to those with the Single-Pixel approach (Figure 4, and dotted lines here). It is evident that the smoothing yields more candidates by a factor of 5-8 for each combination, but the increase in the fraction of random interlopers (red lines) is even more dramatic.}
\end{figure}

The overlap fraction between candidates found in both the single-pixel approach and the search after smoothing may appear on first glance low. Specifically, of the 1756 candidate systems from the former method only 543 remain after smoothing (31 \%), while our estimate of the fraction of real absorbers vs. spurious interlopers and noise artifacts indicates that 69 percent of the the candidates should show a true OVI profile and hence extend over more than one pixel, i.e. are suitable for a successful resolution detection.  In the following, we investigate how these two seemingly contradicting numbers can be reconciled. 

While binning the spectra to the instrumental resolution of SDSS, i.e. combining the flux from 2-3 adjacent pixels, certainly reduces the noise, but since the absorption in neighbouring pixels is typically weaker one may actually lower the signal-to-noise ratio of the candidate. A priori, it is difficult to assess how the detectability of lines is affected by this modification of the search routine. Hence, we have resorted to a Monte-Carlo analysis, creating new data sets of mock absorbers in resolution space based upon the candidates' specifications obtained in the single pixel search. These new sets of candidates were scrutinised by the same algorithm we used to search for absorbers in the real, smoothed spectra.

The recipe for creating one such new data set is as follows : we begin by extracting for each candidate the relevant flux and noise value at each pixel in question for the search criteria (i.e. OVI 1032 \& 1038, Ly $\alpha, \beta$, and $\gamma${} (when possible)). In order to estimate the expected flux values in adjacent pixels, we simply create a Voigt profile which needs to reproduce the flux in the central pixel and has an appropriate Doppler parameter (fixed to b $\sim$ 14 km/s for the OVI, and b $\sim$ 35 km/s for HI). To these expected fluxes in the 2 adjacent pixels, we add noise at the level found for that central pixel. Then we combine the information of the new set of pixels according to the method used for the smoothing of the data, i.e. adding and averaging the fluxes while adding the noise in quadrature only. Hence, we derive for each candidate of the one pixel search a new set of flux and noise values in the relevant transitions. In this way, we construct a new sample of artificially created potential resolution candidates, based upon the assumption that the one-pixel flux values found in our search represent the depression of flux in the centre of the line. The new sample is then subjected to the same search algorithm we used for the real rebinned spectra, and the fraction of candidates passing the search criteria is noted. We repeat this procedure for 10,000 versions of such mock lists, in order to estimate the expectation value and its scatter for the number of candidates remaining.  From this simple procedure, we derive an average "retention rate" of 38 $\pm$ 3 percent, only slightly higher than the value for the overlap percentage in the real data sets. It seems plausible that allowing for more complicated scenarios (line centres not exactly at central pixel, correlated noise, range of Doppler parameters for underlying profiles, etc...) the fraction could be pushed even lower, and hence we conclude that it is not a sign of our search algorithms being unreliable when finding a low overlap percentage.

\end{document}